
\input harvmac
\def\winf{\ W_{1+\infty}\ }
\Title{\vbox{\baselineskip12pt\hbox{MPI-PhT/94-35}\hbox{DFTT-16/94}}}
{\vbox{\centerline{$W_{1+\infty}$ Dynamics of Edge Excitations}
\vskip2pt\centerline{in the Quantum Hall Effect}}}
\centerline{Andrea Cappelli}
\centerline{I.N.F.N and Dipartimento di Fisica, Universit\`a di Firenze}
\centerline{Largo E. Fermi 2, I-50125 Firenze, Italy}
\bigskip
\centerline{Carlo A. Trugenberger}
\centerline{Max-Planck-Institut f\"ur Physik}
\centerline{F\"ohringer Ring 6, D-80805 M\"unchen, Germany}
\bigskip
\centerline{Guillermo R. Zemba}
\centerline{I.N.F.N. and Dipartimento di Fisica Teorica,
Universit\`a di Torino}
\centerline{via P. Giuria 1, I-10125 Torino, Italy}

\vskip .3in
\noindent
Quantum Hall universality classes can be classified by $W_{1+\infty}$
symmetry. We show that this symmetry also governs the dynamics
of quantum edge excitations. The Hamiltonian of interacting
electrons in the fully-filled first Landau level is expressed in terms
of $\winf$ generators. The spectra for both the Coulomb and generic
short-range interactions are thus found algebraically.
We prove the one-dimensional bosonization of edge excitations in the
limit of large number of particles. Moreover, the subleading corrections
are given by the higher-spin $\winf$ generators, which measure the radial
fluctuations of the electron density. The resulting spectrum for the
Coulomb interaction contains a logarithmic enhancement,
in agreement with experimental observations. The spectrum for generic
short-range interactions is subleading and reproduces the classical
capillary frequencies. These results are also extended to fractional
filling by using symmetry arguments.

\bigskip
\bigskip
\Date{July 1994}


\newsec{Introduction}

In a recent series of papers
\ref\ctz{A. Cappelli, C. A. Trugenberger and G. R. Zemba,
{\it Nucl. Phys.} {\bf B 396} (1993) 465.}
\ref\cdtz{A. Cappelli, G. V. Dunne, C. A.
Trugenberger and G. R. Zemba, {\it Nucl. Phys.} {\bf B 398} (1993) 531;
{\it Nucl. Phys. (Proc. Suppl)} {\bf 33 C} (1993) 21.}
\ref\ctza{A. Cappelli, C. A. Trugenberger and G. R. Zemba,
{\it Phys. Lett.} {\bf 306 B} (1993) 100.}
\ref\ctzb{A. Cappelli, C. A. Trugenberger and G. R. Zemba,
{\it Phys. Rev. Lett.} {\bf 72} (1994) 1902.},
we have studied the many-body states at the
{\it plateaus} of the quantum Hall effect
\ref\prg{For a review see: R. E. Prange and S. M. Girvin eds.,
{\it ``The Quantum Hall effect''}, Springer, New York (1990).}
by means of an {\it effective field theory} with
$\winf$ {\it dynamical symmetry}.

The effective field theory approach, developed by Landau, Ginsburg,
Wilson and others \ref\polch{See, for example: J. Polchinski,
{\it ``Effective field theory and the Fermi surface''}, Lectures
presented at TASI-92, Boulder (U.S.A.), June 1992, preprint
NSF-ITP-92-132, UTTG-20-92, hep-th/9210046.}, does not attempt to solve
the microscopic many-body dynamics, but rather it guesses the
macroscopic physics generated by this dynamics.
The variables of the effective field theory are the relevant low-energy
(long-distance) degrees of freedom, which are characterized by a specific
symmetry.
They describe {\it universal} properties, which are independent of the
microscopic details.
This approach is well suited for the quantum Hall effect, where
experiments yield {\it very precise} and {\it universal} values of the
Hall conductivity.

The main input for our effective field theory is the Laughlin theory
of {\it incompressible quantum fluids}
\ref\lau{R. B. Laughlin, {\it Phys. Rev. Lett.} {\bf 50} (1983) 1395;
{\it ``Elementary Theory: the Incompressible Quantum Fluid''}, in \prg ;
{\it ``Fractional statistics in the Quantum Hall effect''},
in F. Wilczek ed. , {\it ``Fractional statistics and anyon
superconductivity''}, World Scientific, Singapore (1990).};
this predicts that the many-body ground states at the plateaus
have uniform density and a gap for density waves. We have shown
that planar incompressible fluids are completely characterized by a
{\it dynamical symmetry} \ref\wyb{See for example: B. Wybourne,
{\it ``Classical Groups for Physicists''},
John Wiley \& Sons, New York (1974), and references therein.}.
This means two basic kinematical properties:

i) each state of the system can be mapped into any other one by a
symmetry transformation, such that the classical configuration space (or
the quantum Hilbert space) carries a representation of the
classical (quantum) symmetry algebra;

ii) the conserved quantities (quantum numbers) characterizing the
excitations are given by a set of symmetry generators, which are
simultaneously diagonal
(the Cartan subalgebra); their spectrum is found by algebraic means.

The standard example of effective field theory with dynamical symmetry is
{\it conformal field theory} in $(1+1)$ dimensions
\ref\gin{For a review see:
P. Ginsparg, {\it ``Applied Conformal Field Theory''}, in {\it ``Fields,
Strings and Critical Phenomena''}, E. Br\'ezin and J. Zinn-Justin eds.,
North-Holland, Amsterdam (1990).} , which
describes  the long-distance physics of critical
phenomena in two space dimensions, and also non-Fermi liquid behaviour of
interacting fermions in $(1+1)$ dimensions \ref\mah{For a review see:
G. D. Mahan, {\it ``Many-Particle Physics''}, Plenum Press,
New York (1990).}.

In the quantum Hall effect, the study of representations of the
$\winf$ dynamical symmetry has given
both the relevant degrees of freedom and their quantum numbers.
We have thus obtained an {\it algebraic} description \cdtz\
of {\it anyons}, collective excitations of
{\it fractional charge, spin and statistics} \ref\wilc{
For a review see: F. Wilczek ed. ,
{\it ``Fractional statistics and anyon
superconductivity''}, World Scientific, Singapore (1990).}, and a
classification of quantum Hall universality classes \ctzb ,
which corresponds to the known {\it hierarchies}
\ref\hiera{F. D. M. Haldane, {\it Phys. Rev. Lett.} {\bf 51} (1983)
605; B. I. Halperin, {\it Helv. Phys. Acta} {\bf 56} (1983) 75 and
{\it Phys. Rev. Lett.} {\bf 52} (1984) 1583; N. Read, {\it Phys.
Rev. Lett.} {\bf 65 } (1990) 1502; B. Block and X.-G. Wen,
{\it Phys. Rev.} {\bf B 42} (1990) 8133 and {\it ibid.} {\bf B 43}
(1991) 8337; for an unified approach, see:
J. Fr\"ohlich and A. Zee, {\it Nucl. Phys.} {\bf 364 B}
(1991) 517; X.-G. Wen and A. Zee, {\it Phys. Rev.} {\bf B 46} (1993)
2290.}.
This exact algebraic description accounts for the precision  of the
measured Hall conductivities at the plateaus.

These results on the kinematical data of incompressible fluids are
reviewed in section 2, setting the language for the new results
discussed afterwards. Let us briefly summarize them.
At the classical level, which is relevant for large samples,
the incompressible fluid is characterized by the dynamical symmetry under
{\it area-preserving diffeomorphisms}, obeying the $w_\infty$ algebra
\ref\she{For a review see: X. Shen, {\it Int. J. Mod. Phys.} {\bf A 7}
(1992) 6953.} .
This is because all possible configurations of a {\it droplet} of
uniform density can be obtained by deformations which preserve the area.
Furthermore, the relevant degrees of freedom are given by the small
fluctuations of a large droplet, say of the shape of a disk, which are
localized at the edge - the so-called {\it edge} waves
\ref\wen{For a review see: X.-G. Wen, {\it Int. J. Mod. Phys.} {\bf B 6}
(1992) 1711.} .
The quantization of these waves yields a $(1+1)$-dimensional
effective field theory, whose Hilbert space consists of a set
of unitary, irreducible, highest-weight representations of
$W_{1+\infty }$, the quantum version of the $w_{\infty }$ algebra \she .

The $W_{1+\infty }$ algebra is linear and infinite dimensional,
like the Virasoro conformal algebra. It involves an infinite
number of conserved currents $V^i_n$, having any positive
(integer) conformal spin $(i+1)$.
Its representations are characterized by the quantum numbers
corresponding to the eigenvalues of $V^i_0$. The basic quantum numbers
are the electric charge $V^0_0$ and the spin $V^1_0$ of the excitations.
The corresponding
currents $V^0_n$ and $V^1_n$ generate the Abelian Kac-Moody and
Virasoro subalgebras of $W_{1+\infty }$, respectively \gin.
The basic operator is not the Hamiltonian, but the Kac-Moody current on
the edge, out of which all the Hilbert space can be constructed.
Therefore, we can identify effective theories of edge excitations
without specifying completely the dynamics of the incompressible fluid.
Moreover, the kinematical data are sufficient to explain the Hall
conduction experiments \cdtz\ctzb\hiera.

In this paper, we incorporate the dynamics
of quantum edge excitations into the framework of $\winf$ symmetry.
Dynamics is important in other types of experiments, where excitations
are induced by temperature or other probes. Here, the effective field
theory is still useful to predict the thermodynamics and the dynamical
response.

In section 3, we discuss the Hamiltonian of edge excitations.
We first show how to use the $w_\infty$ algebra to describe the
dynamics of the classical, chiral, incompressible fluid, and
rederive the {\it capillary frequencies} of its eigenoscillations
\ref\lan{See for example: L. D. Landau and E. M. Lifshitz,
{\it ``Fluid Mechanics''}, Pergamon Press, Oxford (1987).}.
Next, we study the general microscopic quantum
Hamiltonian for interacting electrons in
the fully filled Landau level (filling fraction $\nu =1$).
It contains two competing contributions: the two-body repulsive
interaction and the one-body confining potential,
which holds the electrons together in the droplet.
For large droplets of radius $R\to\infty$, this microscopic dynamics
produces an effective dynamics
for the edge excitations, which can be entirely expressed in terms of
the generators of the $\winf$ algebra, owing to the dynamical symmetry.
The confining potential expands linearly in the generators
$V^i_0$ of the Cartan subalgebra, while the two-body interaction is
represented quadratically in the $V^i_k$.
Both pieces in the effective Hamiltonian are infinite series, which
can be organized according to a natural expansion in powers of $1/R$.
Only a few leading terms are retained in the thermodynamic limit
$R\to \infty $. Specifically, the leading contribution to the
interaction term is shown to be quadratic
in the Kac-Moody generators $V^0_n$, for a large variety of
microscopic interactions.
This result is the proof of the one-dimensional {\it bosonization}
of edge excitations, which has been often used in the literature
\ref\stone{M. Stone, {\it Ann. Phys.} (NY) {\bf 207} (1991) 38,
{\it Phys. Rev.} {\bf B42} (1990) 8399,
{\it Int. J. Mod. Phys.} {\bf B 5} (1991) 509.}\wen ,
but has only been confirmed by numerical studies so far
\ref\sto{M. Stone, H. W. Wyld and
R. L. Schult, {\it Phys. Rev.} {\bf 45} (1992) 14156;
M. Stone and H. W. Wyld, {\it Prog. Theor. Phys.(Suppl.)} {\bf 107}
(1992) 205.}.
Moreover, the $\winf$ algebra describes the subleading
corrections and the associated physical picture.

In section 4, we use the $W_{1+\infty}$ algebra to diagonalize the
Hamiltonian derived in section 3, and compute the spectrum
of quantum edge excitations.
The confining potential always contributes to the leading order $O(1/R)$,
yielding a scale invariant spectrum
${\cal E} (q)= vq$ (with $q\equiv \Delta J/R$ the edge momentum) \cdtz.
The interactions can contribute to the same order and modify the
spectrum, or, otherwise, be negligible.
The {\it long-range} interactions are of the first type.
The {\it short-range} interactions, with range of the order of the
magnetic length $\ell\ $ (the ultra-violet cutoff), belong to the second class.

We prove that the three-dimensional Coulomb interaction
$V(\vert{\bf x-y}\vert ) = e^2 /\vert {\bf x-y}\vert $
modifies the spectrum into ${\cal E}(q)\propto q \log(q)$,
due to the appearance of an infra-red logarithmic singularity.
This result was anticipated by several analytic and numerical
approximations
\ref\volkov{V. A. Volkov and S. A. Milkhailov, {\it Sov. Phys. J.E.T.P.},
{\bf 67} (1989) 1639.}
\ref\wep{X.-G. Wen, {\it Phys. Rev.} {\bf  B 44} (1991) 5708.}
\ref\gps{S. Giovanazzi, L. Pitaevskii and S. Stringari, {\it Phys. Rev.
Lett.} {\bf 72} (1994) 3230.} \sto .
Although this spectrum is no more scale invariant, we stress that it
can be still described by conformal field theory techniques: the
kinematical $\winf$ structure of the Hilbert space of edge excitations
can in fact be used for diagonalizing the Hamiltonian.

Next, we discuss the contribution to the spectrum given by a generic
short-range interaction, with Gaussian fall-off for $|{\bf x}|\gg \ell$.
This contribution has the same form of the classical capillary frequency
- a further indication of the semiclassical nature of the
incompressible fluid \ctza.
The capillary frequency depends on one
phenomenological parameter, which represents the {\it edge tension} of
the one-dimensional boundary. This parameter is found to be
a particular moment of the generic short-range interaction.

The capillary energy, of order $O(1/R^3)$,
is subdominant with respect to the leading contribution of the
confining potential and Coulomb interaction.
This shows that the spectrum of edge excitations is
{\it universal}, {\it i.e.}, independent of short-range effects.
Universality implies experimental predictions that are
{\it parameter-free} \cdtz .

We conclude section 4 by extending our results to the
Laughlin incompressible fluids with fractional filling
$\nu\ =\ 1/3\ ,\ 1/5\ , \dots$.
These are described by other representations of $\winf$
\cdtz\ctzb, and a Hamiltonian which is {\it form-invariant}.

Finally, in the conclusions we discuss the experiments that confirm
the spectrum of edge excitations in the presence of Coulomb
interactions
\ref\weiss{M. Wassermeier, J. Oshinowo, J. P. Kotthaus, A. H.
MacDonald, C. T. Foxon and J. J. Harris, {\it Phys. Rev.} {\bf B 41}
(1990) 10287.}
\ref\ash{R. C. Ashoori, H. Stormer, L. Pfeiffer, K. Baldwin and
K. West, {\it Phys. Rev.} {\bf B 45} (1992) 3894.}.


\newsec{Dynamical symmetry and kinematics of incompressible fluids}

In this section, we shall review the dynamical symmetry of
{\it chiral}, two-dimensional, incompressible fluids and indicate
how this leads uniquely to the construction of the Hilbert spaces
of edge excitations. Next, we shall use their kinematical data
to classify quantum Hall universality classes,
without specifying the form of the Hamiltonian.

\subsec{Classical fluids}

A classical incompressible fluid is defined by its distribution function
\eqn\dif{
\rho (z, \bar z, t) = \rho_0 \ \chi _{S_A(t)}\ ,\qquad
\rho_0 \equiv {N\over A}\ ,}
where $\chi_{S_A(t)}$ is the characteristic function for a surface
$S_A(t)$ of area $A$, and $z=x+iy$, $\bar z= x-iy$ are complex
coordinates on the plane.
Since the particle number $N$ and the average
density $\rho_0$ are constant, the area $A$ is also {\it constant}.
The only possible change in response to external forces is in the
shape of the surface.
The shape changes at constant area can be generated by
{\it area-preserving diffeomorphisms} of the two-dimensional plane.
Thus, the configuration space of a classical incompressible fluid
can be generated by applying these transformations to a reference
droplet.

Next we recall the Liouville theorem,
which states that canonical transformations preserve the
phase-space volume. Area-preserving diffeomorphisms are, therefore,
canonical transformations of a two-dimensional phase space.
In order to use the formalism of canonical transformations, we
treat the original coordinate plane as a {\it phase space}, by
postulating non-vanishing Poisson brackets between $z$ and $\bar z$.
We do this by defining the dimensionless Poisson brackets
\eqn\pob{
\{f, g\} \equiv {i\over \rho_0} \ \left( \partial f
\bar \partial g - \bar \partial f \partial g \right) \ ,}
where $\partial \equiv \partial / \partial z\ $ and
$\ \bar \partial \equiv \partial /\partial \bar z$, so that
\eqn\cmo{\{ z, \bar z\}= {i\over \rho_0}\ .}
Note that the Poisson brackets select a preferred {\it chirality},
because they are
not invariant under the two-dimensional parity transformation
$z\to \bar z,\ \bar z\to z$; in the
quantum Hall effect, the parity breaking
is due to the external magnetic field.

Area-preserving diffeomorphisms, {\it i.e.}, canonical transformations,
are usually defined in terms of a generating function
${\cal L}(z,\bar z)$ of both ``coordinate'' and ``momentum'',
as follows:
\eqn\wdef{
\delta z\ =\ \{{\cal L}, z\}\ ,\qquad
\delta {\bar z}\ =\ \{{\cal L}, {\bar z}\}\ .}
A basis of (dimensionless) generators is given by
\eqn\gen{{\cal L}^{(cl)}_{n,m}\equiv \rho_0 ^{{n+m}\over 2} \ z^n
\bar z^m \ .}
These satisfy the classical $w_{\infty }$ algebra \she
\eqn\clw{\left\{ {\cal L}^{(cl)}_{n,m}, {\cal L}^{(cl)}_{k,l} \right\}
=-i\ (mk-nl)\ {\cal L}^{(cl)}_{n+k-1,m+l-1} \ .}

Let us now discuss how $w_{\infty }$ transformations can be used
to generate the configuration space of classical excitations
above the ground state.
These configurations have a classical energy due to the inter-particle
interaction and the external confining potential, whose specific form
is not needed here.
Let us assume a generic convex and rotation-invariant energy function,
such that the minimal energy configuration $\rho_{GS}$
has the shape of a disk of radius $R$:
\eqn\gsd{
\rho_{GS}(z, \bar z)=\rho _0 \ \Theta \left( R^2-z\bar z
\right) \ ,}
where $\Theta$ is the Heaviside step function.
The classical "small excitations" around this ground state
configuration are given by the infinitesimal deformations
of $\rho_{GS}$ under area-preserving diffeomorphisms,
\eqn\sex{\delta \rho _{n,m} \equiv \left\{ {\cal L}^{(cl)}_{n,m},
\rho _{GS} \right\} \ .}
Using the Poisson brackets \pob , we obtain
\eqn\wav{
\delta \rho _{n,m} = i \left( \rho _0 R^2 \right) ^{{n+m}\over 2}
(m-n)\ {\rm e}^{i(n-m)\theta }\ \delta \left( R^2 -z\bar z\right) \ .}
These correspond to density fluctuations localized on the
sharp boundary (parametrized by the angle $\theta $) of the classical
droplet. Due to the dynamics provided by the energy
function,  they will propagate on the boundary
with a frequency $\omega _k$ dependent on the angular momentum
$k\equiv (n-m)$, thereby turning into {\it edge waves}.
These are the eigenoscillations of the classical incompressible
fluid.

Another type of excitations are classical vortices in the bulk of
the droplet, which correspond to localized holes or dips in the density.
The absence of density waves, due to incompressibility, implies that
any localized density excess or defect is transmitted completely
to the boundary, where it is seen as a further edge deformation.
For each given vorticity in the bulk, we can then construct the
corresponding
basis of edge waves in a fashion analogous to \sex . Thus, the
configuration space of the excitations of a classical incompressible
fluid (of a given vorticity) is spanned by infinitesimal $w_{\infty }$
transformations. This is the property of {\it dynamical symmetry}
described in the introduction.

\subsec{Quantum theory at the edge}

The quantum \footnote{*}{
Throughout this paper we shall use units such that $c=1$, $\hbar =1$.}
version of the chiral, incompressible fluids is given by the
Laughlin theory of the plateaus of the quantum Hall effect \lau .
The simplest example of such a macroscopic quantum state is the
fully filled Landau level (filling fraction $\nu =1$).
Generically, it possesses three types of excitations.
First, there are {\it gapless edge excitations},
which are the quantum descendants of the classical
edge waves described before. These
are particle-hole excitations accross the Fermi surface represented
by the edge of the droplet; therefore, they are called {\it neutral}.
They are also {\it gapless} because their energy, of $O(1/R)$,
vanishes for $R\to\infty$
\wen .
Second, there are localized
quasi-particle and quasi-hole excitations, which have a finite gap. These
are the quantum analogs of the classical vortices and correspond to the
anyon excitations with fractional charge, spin and statistics \lau .
As in the classical case, they manifest themselves as charged excitations
at the edge, owing to incompressibility.
The third type of excitations are two-dimensional density waves in the
bulk,
the magnetoplasmons and (for $\nu <1$) the magnetophonons
\prg
\ref\platz{S. M. Girvin, A. H. MacDonald and P. M. Platzman,
{\it Phys. Rev.} {\bf B 33 } (1986) 2206.}.
These have higher gaps and are not included in our
effective field theory approach.

In the previous section, we have explained the connection between the
classical edge waves and the generators of the algebra $w_{\infty }$
of area-preserving diffeomorphisms.
In the quantum theory, there is a corresponding relation
between edge excitations and
the generators of the quantum version of $w_{\infty }$, called
$W_{1+\infty }$ \she .
This algebra is obtained by replacing the Poisson brackets
\pob \ with quantum commutators: $i\{\ ,\ \}\ \to\ [\ ,\ ]\ $,
and by taking the thermodynamic limit \cdtz .

In this limit, the radius of the droplet grows as
$R\propto \ell\sqrt{N}$, where $\ell=\sqrt{2/(eB)}$ is the
magnetic length and $B$ the magnetic field.
Quantum edge excitations, instead, are confined
to a boundary annulus of finite size $O(\ell )$. In order to construct
an effective field theory for these low-lying excitations, the
quantization  should be followed by the limit
of second-quantized operators into finite expressions localized on
the boundary\footnote{*}{
We refer to our previous papers for a complete discussion of
quantization and the limit of the second-quantized operators to the
boundary \cdtz \ctzb .}.
Let us consider, for example, the case of the fully-filled first Landau
level.
In ref. \cdtz , it was shown that the boundary limit amounts to
a straightforward expansion in powers of $1/R\propto 1/\ell \sqrt{N}$.
The {\it quantum} incompressible fluid, which corresponds
to a Fermi sea of electrons,
becomes effectively a Dirac sea.
Correspondingly, the $(2+1)$-dimensional electrons on the boundary
become Weyl fermions, {\it i.e.}, relativistic, charged, chiral
fermions in $(1+1)$-dimensions \gin .
The field operator of electrons becomes\footnote{**}{
Hereafter, we choose units such that $\ell = 1$.} \cdtz,
\eqn\weyl{
F_R(\theta)\ =\ {1\over \sqrt{ R}}\
\sum_{k=-\infty}^{\infty}\ {\rm e}^{i(k-1/2)\theta}\ b_k\ ,
\qquad \left( |z|=R\ ,\ t=0\right)\ ,}
where $\theta$ parametrizes the circular boundary,
$b_k$ and $b^{\dag }_k$ are fermionic Fock space operators satisfying
$\left\{ b_l, b^{\dag }_k \right\} = \delta _{l,k} \ $,
and $k$ is the angular momentum measured with respect to the ground
state value.

The generators of the quantum algebra $\winf$ are represented
in this Fock space by the bilinears
\eqn\acf{\eqalign{
V_n^j\ &=\ \int_0^{2\pi} {d\theta \over{2\pi}}\ :\
F^{\dagger}(\theta)\ {\rm e}^{-in\theta}\ g^j_n \left(i\partial_{\theta}
\right)\ F(\theta)\ : \cr
\ &=\ \sum_{k=-\infty}^{\infty}\ p(k,n,j)\ : b^{\dag}_{k-n}\ b_k : \ ,
\qquad j\ge 0\ . \cr}}
In this expression,
$F(\theta) = F_R (\theta)\ {\rm e}^{i\theta /2}\sqrt{R}$
is the canonical form of the Weyl field operator of conformal field theory.
The factor $\ g^j_n \left(i\partial_{\theta}\right)\ $
is a $j$-th order polynomial in $i\partial_{\theta}$, whose form
specifies the basis of operators and guarantees the hermiticity
$\left( V^j_n \right)^{\dag} = V^j_{-n}$. The
coefficients $p(k,n,j)$ are also $j$-th order polynomials in $k$ to be
specified later (see Appendix A).
The $\winf$ algebra reads
\eqn\Win{\left[\ V^i_n, V^j_m\ \right] =(jn-im) \ V^{i+j-1}_{n+m} +
q(i,j,m,n) \ V^{i+j-3}_{n+m} +\dots +  c^i(n) \ \delta ^{i,j}
\delta _{n+m,0} \ .}
Here, $i+1=h \ge 1$ represents the ``conformal spin'' of the generator
$V^i_n$, while $-\infty <n< +\infty $ is the angular momentum
(the Fourier mode on the circle).
The first term on the right-hand-side of \Win \ reproduces the classical
$w_{\infty }$ algebra \clw \ by the correspondence
${\cal L}^{(cl)}_{i-n,i} \to V^i_n\ $ and identifies $W_{1+\infty}$ as
the algebra of ``quantum area-preserving diffeomorphisms''.
The additional terms are quantum operator corrections with polynomial
coefficients $q(i,j,n,m)$, due to the algebra of higher derivatives \she.
Moreover, the $c$-number term $c^i(n)$
is the quantum {\it anomaly}, a relativistic effect due to the
renormalization of operators acting on the infinite Dirac sea.
It is diagonal in the spin indices for our choice of basis for the
$\ f^i_k\ $ (see Appendix A).
Finally, the normal ordering ($\ :\ :\ $) of the Fock operators takes
care of the renormalization \gin .

Let us analyse the generators $V^0_n$ and $V^1_n$ of lowest conformal
spin.
{}From \acf \ we see that the $V^0_n$ are Fourier modes of the fermion
density evaluated at the edge $\vert z\vert =R$;
thus, $V^0_0$ measures the edge charge.
Instead, the $V^1_n$ are vector fields which generate angular
momentum transformations on the edge, such that $V^1_0$ measures
the angular momentum of edge excitations.
Their algebra is given by
\eqn\aff{
\left[\ V^0_n, V^0_m\ \right] = c\ n \ \delta_{n+m,0}\ \qquad\qquad,}
and
\eqn\Vir{\eqalign{
\left[\ V^1_n, V^0_m\ \right] &= -m \ V^0_{n+m}\ ,\cr
\left[\ V^1_n, V^1_m\ \right] &= (n-m) \ V^1_{n+m} + {c\over 12}\left(
n^3-n\right) \delta _{n+m,0}\ ,\cr}}
with $c=1$.
These equations show that the $V^0_n$ and $V^1_n$ operators satisfy
an Abelian Kac-Moody algebra and a Virasoro algebra, respectively \gin.

Besides this explicit example for $\nu=1$, leading to a theory
with $c=1$, it has been shown in general that the algebra \Win\
is the unique quantization of the $w_\infty$ algebra in the
$(1+1)$-dimensional field theory on the circle \ref\rava{
I. Vaysburd and A. Radul, {\it Phys. Lett.} {\bf B 274} (1992) 317.}.
For unitary theories, the Virasoro central charge $c$ can be any
{\it positive integer} \ref\kac{ V. Kac and A. Radul, {\it Comm.
Math. Phys.} {\bf 157} (1993) 429.} .
Moreover, there is a one-to-one correspondence between unitary
irreducible representations of the Kac-Moody algebra \aff\ and those
of the $\winf$ algebra \Win \footnote{*}{Apart from special cases,
not relevant here \ref\fren{E. Frenkel, V. Kac, A. Radul and W. Wang,
MIT mathematics preprint, hep-th/9405121.} \ref\awat{H. Awata, M. Fukuma,
Y. Matsuo and S. Odake, preprints YITP/K-1054 and YITP/K-1060,
hep-th/9402001 and hep-th/9405093.} .}.
This means that the operators $V^0_n$, representing the
charge modes on the edge, are the basic building blocks of
the Hilbert space.

Following the standard procedure of two-dimensional
conformal field theory, we define a $W_{1+\infty}$ theory as the
Hilbert space given by a set
of irreducible, highest-weight representations of the $W_{1+\infty}$
algebra, closed under the {\it fusion rules} for making composite
excitations \gin.
Any representation contains an infinite
number of states, corresponding to all the
neutral excitations above a bottom state, the so-called
{\it highest weight} state.
This can be, for example, the ground-state $\vert\Omega\rangle$
corresponding to the incompressible quantum fluid.
The excitations can be written as
\eqn\neut{|\ k,\ \{ n_1, n_2,\dots , n_s \}\ \rangle\ =\
V^0_{-n_1}\ V^0_{-n_2} \cdots V^0_{-n_s } \vert\ \Omega\ \rangle\ ,\quad
n_1 \ge n_2\ge \cdots \ge n_s > 0\ ,}
while the positive modes $(n_i<0)$ annihilate $\vert\ \Omega\ \rangle$.
Here $k=\sum _j n_j$
is the total angular momentum of the edge excitation.

Moreover, any charged edge excitation, together with its tower of
neutral excitations, also forms an irreducible, highest-weight
representation of $W_{1+\infty}$. The states
in this representation have the same form of \neut, but
the bottom state $\vert Q\rangle$ now represents a quasi-particle inside
the droplet.
The charge and spin of the quasi-particle
are given by the eigenvalues of the operators\footnote{*}{
The minus sign is
due to the fact that $V^0_0$ measures the charge on the edge. Due to
overall charge conservation, the charge of a quasi-particle in the bulk
has the opposite sign of its edge counterpart.} $(-V^0_0)$ and $V^1_0$
respectively:
\eqn\hws{V^0_0 \vert Q\rangle =Q\vert Q\rangle\ , \
V^1_0\vert Q\rangle = J \vert Q\rangle\ .}
Actually, all the operators $V^i_0$  are simultaneously
diagonal and assign other quantum numbers to the quasi-particle,
$\ V^i_0 \vert Q\rangle =m_i(Q)\vert Q\rangle, \ i\ge 2\ $,
which are known polynomials in the charge $Q$ (see Appendix A).
These quantum numbers measure the radial moments of the charge
distribution of a quasi-particle (see eq. (3.18) below);
their fixed functional form indicates the rigidity of density
modulations of the quantum incompressible fluid.

All $\winf$ representations, together with their
quantum numbers, were recently obtained in ref. \kac.
In ref. \ctzb ,
we used this representation theory to construct an
algebraic classification of
universality classes of incompressible quantum fluids, which
was found to encompass the general {\it hierarchy} of
plateaus in the quantum Hall effect \hiera .
The incompressible quantum fluids are characterized
by the following kinematical data:

i) the integer central charge $c=m, m \in {\rm Z}$, of the $\winf$
algebra;

ii) the (fractional) charges $Q$ of quasi-particle excitations;

iii) their (fractional) spin $J$ (corresponding to fractional
statistics $\theta/\pi =2 J$);

iv) the value of the filling fraction $\nu$, given by the Hall
conductivity as $\sigma_{xy}=(e^2/h) \ \nu$.
Here we wish to stress that the $\winf$ dynamical symmetry
characterizes completely these kinematical properties of incompressible
quantum fluids, without reference to a specific Hamiltonian.
Note that fractional statistics is also independent
of the dynamics, being a {\it static} property of correlators on the
circle, {\it i.e.}, in the plane. Moreover,
these informations are sufficient to determine $\sigma_{xy}$, and
thus to describe the Hall conduction experiments. Actually,
the Hall current is computable from the chiral anomaly of the
$(1+1)$-dimensional edge theory \cdtz, which is encoded in
the charge commutator eq. \aff\  (see also \ctzb).

Let us finally mention two more kinematical data of the Hilbert space
which were used extensively by X.-G. Wen and other authors to identify
and characterize the edge field theory \wen \ref\napo{
C. Cristofano, G. Maiella, R. Musto and F. Nicodemi, {\it Phys. Lett.}
{\bf B 262} (1991) 88; {\it Mod. Phys. Lett.} {\bf A 6} (1991) 1779,
2985; {\it ibid.} {\bf A 7} (1992) 2583.}.
These are:

i) the number of neutral edge states, {\it i.e.}, the {\it degeneracy}
of the states
\neut\ at fixed angular momentum $k$, which was used to match
the edge theory to numerical spectra of
interacting electrons in the lowest Landau level;

ii) the {\it topological order}, {\it i.e.},
the degeneracy of the ground state
on a toroidal geometry, which is a consequence of the fusion rules
for Virasoro representations \gin .

In summary, the behaviour of quantum incompressible fluids at long
distances is basically characterized by the above kinematical data,
which are expressions
of the structure of the Hilbert space in terms of representations of the
$\winf$ algebra. These data specify the {\it universality classes} for
quantum Hall experiments, where the dynamics (spectrum) of edge
excitations does not play any relevant role.

The characterization of
the Jain sub-hierarchy \ref\jain{For a review see:
J. K. Jain, {\it Adv. Phys.} {\bf 41} (1992) 105.} ,
which describes the plateaus most prominently observed in the
experiments, remains an open issue.


\newsec{$\winf$ dynamics of edge excitations}

In the previous section, we have restricted ourselves to
{\it kinematical} considerations. In the following, we are going to
discuss the {\it dynamics} of the edge excitations.
We start from the classical problem, the spectrum of eigenfrequencies
$\omega_k$ of eigenoscillations.


\subsec{Classical capillary waves}

The dynamics of the classical incompressible fluid is fixed
by a single parameter $\tau $,
with dimension of action, whose role is to translate between the
purely kinematical, dimensionless Poisson brackets \pob \ and the
actual Poisson brackets (with dimension (1/action))
\eqn\eqm{\partial _t \delta \rho _{n,m}={1\over \tau} \{ \delta
\rho _{n,m}, H \} \ = \ i\omega_k\ \delta\rho_{n,m}\ ,
\qquad k=n-m \ .}
The parameter $\tau $ is not generic, but depends on the specific model
under consideration.
In our case, it will be given by the classical
limit of the Laughlin incompressible fluids.

First we study a Hamiltonian which contains only a
rotational invariant
confining potential $V(|{\bf x}|)$. From eq. \eqm , we obtain
the eigenfrequencies,
\eqn\eif{
\omega_k = {V'(R)\over 2R\tau \rho _0} \ k\ ,}
where the prime denotes differentiation with respect to the radial
variable $r=|{\bf x}|$. Note the sign correlation between $k$ and
$\omega _k$, which means that the edge waves propagate only in one
direction along the one-dimensional edge, {\it i.e.}, they are
{\it chiral}.

For applications to the quantum Hall effect, one
should consider a Hamiltonian also containing the
two-body repulsive interaction.
A known method for handling the interaction at the
classical level is the {\it hydrodynamic approximation} \lan
\ref\dun{We thank Gerald
Dunne for pointing this out to us.}.
The two-body
interaction is replaced by the one-body effect of a {\it boundary
pressure} arising from the imperfect cancellation of the repulsive forces
near the edge of the sample. We expect this approximation to be good for
inter-particle interactions which are not too long ranged,
as in a liquid. This expectation
will be confirmed by the quantum dynamics discussed below.

We thus have the following decomposition of the boundary force:
\eqn\dec{V'(R)=V'_{cp}(R) + V'_p(R)\ .}
The pressure contribution
to the frequencies $\omega _k$ is known as {\it capillary frequency}
\lan, and can be computed as follows. The
radial force due to the interactions is given by $F_p(r)=-p'(r)/
\rho _0$, where $p(r)$ is the pressure. Therefore, we have $V'_p(R)=
p'(R)/\rho _0$.
Let us consider an harmonic
edge oscillation parametrized by $r=R+\epsilon (\theta)$, with
$\epsilon (\theta )=\epsilon _0 {\rm sin}(k\theta )$, and expand
\eqn\expa{p(r)=p(R)+p'(R)\epsilon \ .}
We now use the following representation of the boundary pressure:
\eqn\pre{
p=-{\alpha \over R_c}\ =\ - \alpha\ {|\partial _{\theta} {\bf x}\wedge
\partial ^2_{\theta} {\bf x}| \over |\partial _{\theta } {\bf x}|^3 }
=-\alpha\ {2\left( \partial _{\theta} \epsilon \right) ^2 + r\left( r-
\partial ^2_{\theta } \epsilon \right) \over \left( r^2+ \left(
\partial _{\theta} \epsilon \right) ^2 \right) ^{3\over 2} }\ .}
where $\alpha $ is the {\it edge tension}, which is positive
for repulsive interactions, and $R_c$ represents
the local radius of curvature.
Expanding this expression up to first order in $\epsilon $
we find,
\eqn\efe{-p(R+\epsilon)={\alpha \over R} - {\alpha \over R^2}
\left( \epsilon +\partial ^2_{\theta }\epsilon \right) + O\left(
{\epsilon ^2\over R^3} \right)\ .}
Using $\partial ^2_{\theta }\epsilon =-k^2 \epsilon $, we
obtain
\eqn\ppo{V'_p(R)=-{\alpha \over R^2\rho _0} (k^2-1)\ ,}
and the frequencies
\eqn\ffr{\omega _k= {V'_{cp}(R)\over 2R\tau \rho _0} \ k -
{\alpha \over 2R^3\tau \rho _0^2} \ k (k^2-1)\ .}

As a last step, it remains to determine the value of $\tau $ relevant
to the classical limit of the Laughlin incompressible quantum fluids.
Consider the Lagrangian for planar electrons of mass $m$ in
an external magnetic field $B$,
\eqn\lag{L=\sum_{i=1}^N \left( \ {m\over 2} \dot {\bf x} _i^2\ +\
e {\dot {\bf x}}_i \cdot {\bf A}({\bf x}_i)\ \right)\ \ ,}
in the symmetric gauge ${\bf A}({\bf x})={B\over2}(-y,x)$, with $B > 0$.
The Laughlin fluids involve electrons in the first Landau level only.
Thus, we can project out the higher Landau levels in the limit of large
$B$.
As was shown in \ref\djt{G. V. Dunne, R. Jackiw and
C. A. Trugenberger, {\it Phys. Rev.} {\bf D 41} (1990) 661.} ,
this projection
can be implemented by taking the limit $m\to 0$ of
\lag . The residual dynamics is then governed by the following
Lagrangian:
\eqn\rdy{L_R=\lim _{m\to 0}L = {eB\over 2} \sum _{i=1}^N {\bf x}_i
\wedge \dot {\bf x}_i \ .}
This Lagrangian is of first order in time derivatives, which implies
that the projection on a subspace of constant (kinetic) energy
amounts to a phase space reduction from a 4$N$-dimensional phase-space
to a 2$N$-dimensional phase-space. The new symplectic structure is
given by \djt
\eqn\nsy{\{z_i, \bar z_j\} _{PB}=i {2\over eB} \delta _{ij}\ ,}
and shows that the original coordinate plane of the electrons behaves
indeed as a phase space for motions at constant kinetic energy.
By comparing \nsy \ with \pob \ we obtain the explicit expression
for the parameter $\tau $:
\eqn\pat{\tau={eB\over 2\rho _0}\ .}
Inserting this in \ffr , we obtain the final result for the
dispersion relation of the classical edge waves:
\eqn\fdr{\omega _k={v\over R} k- {\alpha \over eB\rho _0R^3} k
(k^2-1)\ ,\qquad v={V'_{CP}(R)\over eB}\ .}

This formula can be also obtained by solving the hydrodynamic equations
in the presence of a magnetic field \volkov \gps.
Here, we showed that it can be derived
in a much simpler way after realizing that the kinematics of the
incompressible classical fluid is specified by
$w_\infty$ transformations\footnote{*}
{Note, however, that the hydrodynamic equations also give a higher,
gapful branch of frequencies for compression waves corresponding to the
magneto-plasmon.}.

Equation \fdr\ has some properties which are worth stressing and must
be kept in mind for the quantum dynamics. First, there are zero modes,
which correspond to symmetries: the $k=0$ radial breathing mode of the
droplet is clearly absent in an incompressible fluid, as is evident from
\wav.
In addition, a translation invariant
two-body interaction gives rise to a vanishing
capillary frequency for the modes $k=\pm 1 $, which have
$\epsilon (\theta)\propto \sin\theta$ and
thus correspond to an overall translation of the droplet.
Another general property of the classical fluid is the degeneracy of its
eigenoscillations, as it is also apparent from eqs. \wav\ and \eqm :
although a generic $w_\infty$ deformation of the droplet is parametrized
by two integers $(n,m)$, the frequency depends on their difference
$k$ only.
This is the consequence of the sharp boundary of the classical droplet,
which forbids infinitesimal radial fluctuations.
On the contrary, the quantum incompressible fluid has a smooth boundary,
and radial fluctuations produce a hierarchy of dynamical effects,
which are described by the action of the quantum algebra $\winf$.

Finally, we note that $\omega_k$ contains two {\it phenomenological
parameters} $v$ and $\alpha$, in addition to the generic parameters
$\rho_0$, $A=\pi R^2$ and $B$.
For large droplets, the confining potential term dominates over the
capillary
frequency and the dispersion relation becomes of relativistic nature,
with {\it effective light-velocity} $v$.


\subsec{Quantum hamiltonian in the first Landau level}

In the following, we shall discuss the quantum Hamiltonian for the
incompressible fluid at filling $\nu=1$.
We consider the microscopic model of planar electrons
confined to the first Landau level.
This projection quenches the kinetic energy
\ref\girj{S. M. Girvin and T. Jach, {\it Phys. Rev.}
{\bf B 29} (1984) 5617.}, so that the Hamiltonian
contains only the confining potential $V_{CP}(|{\bf x}|)$
and the repulsive two-body interaction $V_I(|{\bf x} -{\bf y}|)$,
\eqn\reh{
H=H_{CP} +H_I=
\int d^2 {\bf x}\ V_{CP}({\bf x})\ \rho ({\bf x}) + {1\over 2}
\int d^2{\bf x} \int d^2{\bf y} \ \rho ({\bf x})\ V_I({\bf x}-{\bf y})\
\rho ({\bf y}) \ ,}
where $\rho ({\bf x}) \equiv \Psi ^{\dag }({\bf x}) \Psi ({\bf x})\  $
is the density operator, constructed from the field operator
\eqn\fop{
\Psi ({\bf x}) =\sum _{j=0}^{\infty} a_j \ \psi _j ({\bf x}) \ .}
Here, $a_j$ and $a^{\dag }_j$ are fermionic Fock space annihilators and
creators, satisfying the usual anticommutation rules
$\ \left\{ a_l, a^{\dag }_k \right\} = \delta _{l,k} \ ,\ $
with all other anticommutators vanishing, and $\psi _j({\bf x})$ are the
first Landau level wave functions
\eqn\flw{\psi _j({\bf x}) = {1\over \ell \sqrt{\pi j!}} \
\left( {z\over \ell}
\right) ^j \ {\rm e}^{-{{|z|^2}\over 2\ell ^2}}\ .}
In this equation, we have restored the magnetic length $\ell$ ,
for the sake of completeness.

In the following, we shall write the Hamiltonian
in terms of quantum $\winf$ operators, which will be used later to
find the spectrum of quantum edge excitations.
The operators producing the quantum analogs of the classical deformations
$\delta_{n,m}\rho$ have been defined in ref. \ctz, by quantizing the
classical generators \gen\ in the first Landau level:
\eqn\lnm{\eqalign{
{\cal L}_{n,m} &= \int d^2 {\bf x}\ \Psi ^{\dag}({\bf x})\ \left(
b^{\dag} \right) ^n \left( b \right) ^m\ \Psi ({\bf x}) \ ,\cr
b &\equiv {\bar z\over 2} + \partial \ ,\qquad
b^{\dag } \equiv {z\over 2} - \bar \partial \ .\cr}}
These operators satisfy a ``non-relativistic'' version of the
quantum algebra $\winf$ \Win\ \ctz\
\ref\iso{S.Iso, D. Karabali and B. Sakita, {\it Nucl. Phys.}
{\bf B 388} (1992) 700; {\it Phys. Lett.} {\bf 296 B} (1992) 143;
I. Kogan, preprint hep-th/9401093.}.
The operators ${\cal L}_{nn}$ are mutually commuting, as in the
classical case,
and can be simultaneously diagonalized. The physical meaning of their
eigenvalues can be inferred from the expectations values
\eqn\rmom{
\langle {\cal L}_{nn} \rangle\ = (-1)^n\ 2\pi\ n!\
\int_0^\infty \ dr\ r\ L_n({r^2})\ \langle\rho(r^2)\rangle\ ,}
where the $L_n(x)$ are the Laguerre polynomials
\ref\gradst{I. S. Grashteyn and I. M. Ryzhik, {\it ``Table of
Integrals, Series and Products''}, Academic Press, London (1980).}.
Therefore, they measure certain {\it radial moments}
of the density distribution in any given quantum state.
In order to describe quantum numbers of excitations, they have to be
normal ordered by subtracting the ground-state expectation values,
${\cal L}_{nn}\to \ :{\cal L}_{nn}:={\cal L}_{nn}-
\langle\Omega |{\cal L}_{nn}|\Omega\rangle$,
where $|\Omega\rangle$ is the filled Fermi sea
\eqn\grst{
|\Omega\rangle=a^{\dagger}_0 \cdots a^{\dagger}_N |0\rangle\ ,}
of $(N+1)$ electrons. Given that
$\langle\Omega |{\cal L}_{nn}|\Omega\rangle = O(N^{n+1})$,
the ${\cal L}_{nm}$ operators, as well as their ``non-relativistic''
algebra, undergo a severe renormalization in the large $N$ limit.
It can be shown that specific linear combinations
$\sum_{n=0}^i\ \lambda_n \ {\cal L}_{n-k,n}\ $
have a finite limit as the normal-ordered  boundary operators $V^i_k$
introduced in \acf, satisfying the $\winf$ algebra \Win.
Therefore, we shall expand the Hamiltonian in the basis of the
operators $V^i_k$.

Let us first consider the one-body term $H_{CP}$.
As discussed in ref. \cdtz, for $\nu =1$ the geometrical
(semi)-classical picture of edge waves has a simple correspondence in
the Fock space of fermions. Edge waves are particle-hole excitations
above the filled Fermi sea \grst\
with angular momentum $k=\Delta J \sim O(1)$, $|k|\ll \sqrt{N}$ \cdtz.
The boundary field theory is constructed by operators which measure these
transitions and remain finite for large $N$.
Let us consider the Fock expression of $H_{CP}$ and expand the
confining potential in a power series:
\eqn\potser{
V_{CP}(|{\bf x}|)= \gamma_0 +\gamma_1 |{\bf x}|^2 +\gamma_2 |{\bf x}|^4
+ \cdots\ ,}
\eqn\hcp{
H_{CP}=\sum_{k=0}^\infty \left[ \gamma_0 +\gamma_1\left(k+1\right)
+\gamma_2\left(k^2 + 3k +2\right)+\cdots \right]\ a^\dagger_k a_k\ .}
Following ref. \cdtz, it is convenient to redefine the Fock operators
by shifting the angular momentum index,
\eqn\bfo{
b_r\ =\ a_{N+r}\ ,\qquad\qquad b^{\dag}_s\ =\ a^{\dag}_{N+s}\ ,}
such that the first empty state has index $r=1$.
For large $N$, the excitations of electrons from the deep inside of the
droplet have high energy and can be neglected. Thus, we can extend
the summation in \hcp\ from $k\in[-N,\infty)$ to
$k\in(-\infty, +\infty)$, {\it i.e.}, consider a relativistic Dirac sea.
We can now compare $H_{CP}$ in \hcp\ to the explicit Fock space
expression of the $\winf$ operators \acf\ (see ref. \cdtz\ and
appendix A),
\eqn\vofs{\eqalign{
V^0_n\ &=\ \sum_{r=-\infty}^{\infty}\ :b^{\dag}_{r-n} b_{r}:\ , \cr
V^1_n\ &=\ \sum_{r=-\infty}^{\infty}\
\left( r- {n+1 \over 2}\right)\ :b^{\dag}_{r-n} b_r:\ , \cr
V^2_n\ &=\ \sum_{r=-\infty}^{\infty}\
\left( r^2- (n+1)r +{(n+1)(n+2)\over 6} \right)\ :b^{\dag}_{r-n} b_r:\ ,
\cr}}
and obtain the desired expansion,
\eqn\hcpd{\eqalign{
H_{CP}  -\langle\Omega|H_{CP}|\Omega\rangle\ &=\ \gamma_0 V^0_0 +
    \gamma_1 \left(V^1_0 + \left( N+{3\over 2}\right) V^0_0 \right) \cr
\ &+\ \gamma_2 \left(V^2_0 + 2\left(N+2\right) V^1_0 +
  \left( (N+2)^2-{1\over 3}\right)V^0_0\right) +\cdots\ .\cr}}
In general, confining potentials growing as $V_{CP}=O(|{\bf x}|^{2i})$
involve all operators $V^j_0$ with $j\le i$.

We consider the confining potential $V_{CP}$ to be generated dynamically
at the edge of the sample.
Therefore, we expect a self-tuning of the parameters $\gamma_i$
for large $N\equiv R^2$,
such that the normal-ordered energy is finite when written in terms
of the quantum numbers $\langle V^i_0\rangle$ of edge excitations.
We thus consider the normal-ordered Hamiltonian
\eqn\hno{
H_{CP}= \alpha_0\ (V_R)^0_0\ +\ \alpha_1\ (V_R)^1_0\ +\
\alpha_2\ (V_R)^2_0\ +\dots\ ,\qquad N\to\infty\ ,}
where the $(V_R)^i_0$ are the generators of $\winf$ properly
normal-ordered on the {\it cylinder} $(R\theta,t)$.
They are related to the $V^i_0$ of eq. \acf , conventionally defined
on the {\it conformal plane}, by well-known transformation rules
\ref\car{For a review see: J. L. Cardy,
{\it ``Conformal Invariance and Statistical Mechanics''}, in
{\it ``Fields, Strings and Critical Phenomena''},
E. Br\'ezin and J. Zinn-Justin eds., North-Holland, Amsterdam (1990).}
(see Appendix A).
The resulting expression is
\eqn\hren{H_{CP}\ =\ \alpha _0 V^0_0+ {\alpha _1\over R}
\left( V^1_0 -{c\over 24}\right) +
{\alpha _2\over R^2} \left( V^2_0-{1\over 12} V^0_0\right)
+ O\left( {1\over R^3} \right), \qquad  R\equiv\sqrt{N}\  .}
Note that the $(V_R)^i_0$ are dimensionful, because their
eigenvalues are of order $O\left( (k/R)^i \right)=O(q^i)$ for states
of angular momentum $k$ and edge momentum $q$, as explained
in section $4$.
A sensible thermodynamic limit of the edge theory is achieved if
the {\it phenomenological} coefficients $\alpha_i$ are asymptotically
{\it independent} of $R$ (modulo logarithmic corrections).
In this case, the higher-spin operators $V^i_0\quad (i\ge 2)$ measure
the deviation from the massless, critical spectrum
${\cal E}_k=vk/R=vq$.


\subsec{Two-body interactions}

The interaction term $H_I$ of the Hamiltonian \reh\ can be treated
similarly. In Fock space, it reads
\eqn\dh{\eqalign{
H_I\ =\ \sum_{n=0}^{\infty} \sum_{m=0}^{\infty}
\sum_{k=-m}^{n}\ {1\over 2} &\left( M(n,m;k)\ -\ M(n,m;n-m-k)\ \right)\
a^{\dag}_{m+k} a^{\dag}_{n-k} a_n a_m \ , \cr
M(n,m;k)\ \equiv {1\over 2}\ \int d^{2} z_1 \int d^2 z_2 &\
{\psi}^{*}_{n-k}(z_1 , {\bar z}_1 )\ \psi _n (z_1 , {\bar z}_1 )\
 V_I(|z_1 - z_2|)\ \cdot \cr
\ & {\psi}^{*}_{m+k}(z_2 , {\bar z}_2 )\
\psi _m (z_2 , {\bar z}_2 )\ ,\cr}}
where we have explicitly anti-symmetrized the matrix element.
We study again $H_I$ in the large $N$ limit for
transitions corresponding to edge excitations, which involve
indices $n=N+r$ and $m=N+s$ with $|r|,|s|, |k| \sim O(1)$. In these
cases, the ``transition density'' describing each virtual electron
in \dh\ can be approximated as follows:
\eqn\aswf{
\rho_{s,k}(z,{\bar z})\equiv{\psi}^{*}_{N+s+k}(z,{\bar z})\
\psi _{N+s} (z,{\bar z})\ \simeq\ {1 \over{\pi\sqrt{2 \pi N}}}\
{\rm e}^{ik \theta}\
\exp\left(-{(|z|^2 - (N +s +{k\over 2}))^2 \over{2N}}\right)\ ,}
where $z= |z| {\rm e}^{i\theta}$.
The phase of this expression is the chiral edge wave of momentum $k$.
The modulus has the shape of a Gaussian, with center located at
$|z| \simeq R + (2s+k)/4R$, and spread $\Delta |z| = O(1)$.
This shows that all virtual electrons participating in edge transitions
are localized on a boundary annulus of the size of the ultra-violet
cut-off.
We can develop some physical intuition on the form of the matrix element
by thinking of the interaction between the densities of the two electrons
through their phases (index $k$) and modulus (radial) fluctuations
(indices $r,s$).

Consider in the first place the (three-dimensional)
Coulomb potential\footnote{*}{We shall take the dielectric constant
equal to one.}
\eqn\Coul{V(|z_1 - z_2|)= {e^2 \over |z_1 - z_2|} \ .}
The two integrals in \dh\ are effectively restricted to the
annular region. One can distinguish two
contributions: when $|z_1-z_2| >1$, {\it i.e.},
the relative angle is $|\theta|=|\theta_1 - \theta_2| > 1/R$, and
its complement.
In the first case, the two densities \aswf\ are far apart, so that
their radial fluctuations of $O(1/R)$ can be neglected.
This allows the further approximation,
\eqn\cwf{
\rho_{s,k} (z,{\bar z})\ \simeq\ {1\over{2\pi R }}\ {\rm e}^{ik\theta}\
\Theta \left( \left(|z| -R\right)\left( R+1-|z| \right) \right )\ .}
Therefore, the matrix element has only one non-trivial integration over
the relative angle:
\eqn\ddy{
M(k)\ \equiv\ M(N+r ,N+s ;k)\ \simeq\ {e^2 \over{2 R}}
\int_{1/R}^{\pi}\ {d\theta\over 2\pi}\
{\cos ( k \theta ) \over \left\vert \sin {\theta\over 2} \right\vert }.}
In the other region, $|z_1-z_2| <1$, the integrals are fully
two-dimensional. Nevertheless, the ultra-violet behaviour of \Coul\ is
integrable and the result is subleading with respect to \ddy.
The integral in \ddy\ can be computed by a recursion relation in $k$,
and reads\footnote{**}{We have also obtained this
result by more accurate two-dimensional integrations.},
\eqn\ddz{
M(k)=M(|k|) \simeq M(0)- {e^2 \over{\pi R}}\ \sum_{l=1}^{k}
{1\over{2l-1}}\ .}

The Coulomb matrix element $M(k)$ exhibits a logarithmic behaviour
for $k\gg 1$,
\eqn\codi{M(k)= M(0) -\ {e^2 \over{2\pi R}}\ \log k\ ,\qquad
\qquad k \gg 1\ ,}
which is accessible within the validity of the edge wave approximation
$k\ll R$ \cdtz.
This behaviour is due to the infrared singularity of the
electrostatic energy of charges in the annular region.
Note that this singularity cannot be screened by the ion background,
which ensures charge neutrality over distances larger than $R$.
\bigskip
For interactions that are not too long-ranged, both phase and radial
fluctuations of $\rho_{s,k} (z, {\bar z})$ could be relevant in \dh,
and give an involved dependence of the matrix element on $(r,s,k)$.
We can establish a general result in the opposite
regime of (very) {\it short-range} interactions, whose range is of the
order of the ultra-violet cut-off. A convenient
way to expand these potentials is given by the Gaussian
basis of Ref. \girj
\eqn\gaus{
V_I(|z_1 - z_2|)\ =\ \int_{0}^{\infty}\ dt\ f(t)\
{\rm e}^{-t|z_1 - z_2|^2}\ .}
For short-range interactions, $f(t)$ has a compact support over
values of $t \simeq O(1)$. The matrix element of the Gaussian
interaction,
\eqn\mesr{\eqalign{
M_t(N+r,N+s;k)\equiv {1\over 2}\ \int d^{2} z_1 \int d^2 z_2 &\
{\psi}^{*}_{N+r-k}(z_1 , {\bar z}_1 )\ \psi _{N+r} (z_1 , {\bar z}_1 )\
 {\rm e}^{-t|z_1-z_2|^2}\ \cdot \cr
\ & {\psi}^{*}_{N+s+k}(z_2 , {\bar z}_2 )\
\psi _{N+s}(z_2 , {\bar z}_2 )\ ,\cr}}
can be evaluated for $N\to\infty$ using the
saddle-point technique, which is equivalent to the approximation in
\aswf\ (for details see Appendix B). The resulting
expression is
\eqn\meres{\eqalign{
M_t(N+r,N+s;k)\ \simeq & {1\over{\sqrt N}} {1\over 4 \sqrt{\pi}
\sqrt{t(1+t)} } \left\{ 1 +{1\over 4N} \left[
-{1\over t}\left( k^2 - {1\over 4} \right) \right.\right. \cr
& +2\left( k(r-s-k)-r^2-s^2-{r+s \over 2}-{5\over 24}\right) \cr
& \left.\left. +{1\over 1+t} \left(
k^2-2k(r-s)+2(r^2 + s^2) + t(r+s)^2 \right)
\right]\right\}\ . \cr}}
One can verify that this expression is
symmetric under the interchange of the two particles,
$(r,s,k) \leftrightarrow (s,r, -k)$.
Moreover, the anti-symmetric matrix element in \dh\ becomes
\eqn\asme{
{1\over 2} \left(  M_t(k)\ -\  M_t(r-s-k)\right) =
- {1 \over R^3 32 \sqrt{\pi} }\ {(1+2t)\over (t(1+t))^{3/2} }\
\left[ k^2\ -\ (r-s-k)^2\ \right] \ . }
We can now insert this result back into the Hamiltonian
\dh\ and exploit again the antisymmetry in the indices
$k$ and $(r-s-k)$, to obtain a matrix element
which depends on $k$ only, as in the case of the Coulomb interaction.
Actually, eq. \asme\ has the unique index structure, at most quadratic in
$(r,s,k)$, which has the above symmetries.

The next step is to rewrite the Hamiltonian in terms of $\winf$
generators, as done for $H_{CP}$ (see eqs. \vofs -\hno). For the
Coulomb and short-range interactions, equation \dh\ takes the form
\eqn\dhs{
H_I\ =\ \sum_{r=-N}^{\infty}\sum_{s=-N}^{\infty} \sum_{k=-N-s}^{N+r}\
{\lambda_k\over 2}\
b^{\dag}_{s+k} b^{\dag}_{r-k} b_r b_s \ ,}
where the $b$'s are the Fock operators with shifted index \bfo, and
$\lambda_k/2$ is the matrix element in the limit of edge excitations,
\ddz\ and \asme .
Due to the explicit antisymmetrization of \dh , however,
these matrix elements are determined up to a $k$-independent additive
term, which vanishes in any expectation value.
This freedom will be useful later.

We manipulate eq. \dhs\ by letting $N\to\infty$ in the summation
extrema and by replacing the four Fermi operators with a pair of
$V^0_k$ operators \vofs.
Moreover, we introduce the normal ordering $\ :\ :\ $ with respect to the
incompressible ground state \grst. This subtracts a one-body operator
(assigned to $H_{CP}$) and a constant from \dhs.
The first subtraction gives,
\eqn\iseh{
H_I\ =\ \sum_{k=-\infty}^{\infty}\ {\lambda_k\over 2}\
:\ V^0_{-k} V^0_k\ :\ .}
The second subtraction requires some knowledge of $\winf$
representations. Let us recall from Section $2.2$, eq. \neut,
that the ground state is the highest weight state of a $\winf$
representation, which satisfies the conditions
$V^i_k |\Omega\rangle =0$ , for $k\ge 0$ and $i\ge 0$ .
Therefore, the normal ordered expression is found to be\footnote{*}{
Note that finite-size effects are absent in this formula, in particular
$(V_R)^0_k =V^0_k$ (see appendix A).}:
\eqn\ieh{
H_I\ =\ {\lambda_0 \over 2}\ \left(V^0_0 \right)^2 \ + \
\sum_{k=1}^{\infty}\ \lambda_k\ V^0_{-k} V^0_k \ .}

An important property of the microscopic Hamiltonian \dh\ is the
translation invariance of two-body interactions.
This is enforced on the edge approximation \ieh\ as a further
normal ordering condition, which fixes the up to now free
additive constant in $\lambda_k$.
As pointed out in section 3.2 , the edge excitations with $k=\pm 1$
correspond to infinitesimal overall translations of the droplet.
At the quantum level, they are generated by $V^0_{\mp 1}\ $.
Given that the translation of any state should cost no energy, we require
\eqn\triv{
[\ H_I\ ,V^0_{-1}\ ]\ =\ 0\ \quad \to \quad\
\lambda_1 \ =\ 0 \ ,\qquad\qquad {\rm (translation\ invariance)}\ .}

In conclusion, the Hamiltonian for edge excitations with the
Coulomb and short-range interactions is given by \ieh , with
\eqn\lafin{\eqalign{
\lambda_k \ =& \ -{2e^2 \over \pi R}\ \sum_{l=2}^k {1\over 2l-1}\
\simeq -{e^2 \over \pi R}\ \log k\ ,\quad
\ \left( \lambda_0 ={2e^2 \over \pi R}\right)\ ,
\qquad {\rm (Coulomb\ interaction)}, \cr
\lambda_k \ =&\ - {k^2 - 1 \over{8 \sqrt{\pi} R^3}}\ \int_0^{\infty} \
dt\ f(t)\ {(1+2t)\over{ (t(1+t))^{3/2}}}\ ,
\qquad\qquad\qquad {\rm (short-range \ interactions)}. \cr}}

Let us make a few remarks on the result \ieh\ for the edge Hamiltonian.
This form for $H_I$ has been assumed in phenomenological and semiclassical
descriptions of the edge excitations \stone\wen\wep:
a one-dimensional {\it chiral boson} $\varphi(R\theta-vt)$,
which expresses the fluctuations of the density at the edge by
$\rho(\theta)=-{1\over 2\pi R}\partial_\theta\varphi(\theta)\ $ \cdtz,
is introduced. Moreover, its Hamiltonian is taken to be
\eqn\hieff{
H_I^{(boson)}\ = \ {1\over 2} \int_{0}^{2\pi} d\theta_1
\int_{0}^{2\pi} d\theta_2
\rho(\theta_1)\ V_I^{(eff)}(|\theta_1-\theta_2|) \ \rho(\theta_2)\ .}
Given that the chiral boson carries a representation of the Kac-Moody
algebra \aff\ , we can express
$\rho(\theta)=\sum_k \ \exp(ik\theta)\ V^0_k$ and recover the operator
form in \ieh.
In this description, as well as in the semiclassical limit,
the radial degree of freedom is completely neglected, thus the
matrix element can only depend on the Fourier mode $k$ on the edge.
This {\it bosonization} of the edge waves in the quantum Hall effect
was predicted by Stone \stone\ and later verified numerically \sto .

Our derivation can be regarded as the first {\it proof}
of the bosonization, for both the Coulomb and generic
short-range interactions.
Moreover, the $\winf$ dynamical symmetry encodes the physical picture
and the subleading corrections.
The bosonization formula \ieh\ and the operators $V^0_k$ describe the
leading semiclassical effect of phase fluctuations at the edge.
Corrections can have the following structure:
equation \rmom\ shows that the $V^i_0, \ i>0,$
measure the radial fluctuations of the density at the edge \aswf\ ,
which appear as subleading corrections in the $1/R$ expansion of the
matrix elements.
These corrections can be expressed in terms of the $V^i_k,\ i>0\ ,$
via their Fock expressions \vofs.
For example, the first subleading correction to \ieh\ has the form
\eqn\hisub{\eqalign{
H_I^{(sub)}\ =&\ \sum_{k>0}\ \eta_k\ \left(
      V^0_{-k} V^1_{k}\ +\ V^1_{-k} V^0_k \right) \cr
\ =&\ \int_{0}^{2\pi} d\theta_1 \int_{0}^{2\pi} d\theta_2
\rho(\theta_1)\ V_I^{(sub)}(|\theta_1-\theta_2|) \
:\rho(\theta_2)^2:\ .\cr}}

The bosonization of edge waves in the presence of the
Coulomb interaction is rather easy to understand, because radial
fluctuations in \aswf\ , of size $O(1/R)$,
are subleading for long-range $O(R)$ interactions.
The bosonization of short-range interactions is, instead, due to
the semiclassical nature of the incompressible fluid. As we show
in the next section, their spectrum actually reproduces the classical
capillary frequencies $\omega_k$ of section 3.1.
Although {\it a priori} radial excitations could be
important at short distances, they are quenched by
incompressibility, which is enforced by Fermi statistics (see eq. \asme ).

In conclusion, the bosonization of edge excitations is a
semiclassical effect.
Here we provided both a physical picture and a technique to derive it
from the microscopic quantum theory.
Moreover, we indicated the generic form and the origin of subleading
corrections to the bosonization formula.


\newsec{The spectrum of edge excitations}

Up to now we have derived the generic form of the Hamiltonian
$H\ =\ H_{CP}\ +\ H_I$ governing the dynamics of quantum edge
excitations for $\nu=1$, and we have shown that it can be written
entirely in terms of $\winf$ generators. The piece $H_{CP}$,
describing the effect of the confining potential, is a {\it linear}
combination of the generators $V^i_0$ of the Cartan subalgebra,
see eq. \hren . The piece $H_I$, describing
the two-body interactions, is quadratic,
see eq. \ieh .

The diagonalization of $H$ follows in principle from the
algebraic properties of the operators $V^i_k$ only, owing to the
dynamical $\winf$ symmetry. In practice, however,
the diagonalization is rendered difficult by the fact that
\eqn\noc{
[\ V^i_0\ ,\ H_I\ ]\ \neq\ 0\ ,\qquad i \ge 2\ .}
This implies that there is no general basis of quantum edge excitations,
which diagonalizes simultaneously all terms of the confining potential
and the interaction. Actually, each contribution to the Hamiltonian is
diagonal in a known basis: the operators
$(V^0_0,V^1_0, H_I)$ are diagonal in the standard {\it bosonic}
basis \neut\ ;
the Cartan subalgebra $(V^i_0,\  i\ge 0)$, is diagonal in the
{\it fermionic} basis to be discussed later.
Therefore, we shall analyse each interaction separately, choosing
the most appropriate basis according to the dominant operators in the
$(1/R)$ expansion.


\subsec{Coulomb interaction}

The Coulomb interaction $V({\bf x})= e^2/|{\bf x}|$ gives rise to an
interacting Hamiltonian of order $O(1/R)$, eqs. \ieh\ \lafin.
In this case, the generators $V^i_0\ ,\ i \ge 2$, in the
expansion of $H_{CP}$ lead to subdominant contributions
to the spectrum for large $R$ and can be dropped.
The only term which must be retained is the leading $1/R$
contribution, which is proportional to $V^1_0$.
We thus consider the Hamiltonian:
\eqn\hcoul{
H^{(Coulomb)}\ =\ \alpha_0 V^0_0 +
{\alpha_1\over R}\left( V^1_0 - {c\over 24} \right) +
\ {\lambda_0 \over 2}\ \left(V^0_0 \right)^2 \ + \
\sum_{k=1}^{\infty}\ \lambda_k\ V^0_{-k} V^0_k \ +
O\left({1\over R^2}\right)\ ,}
where the conformal central charge for $\nu=1$ is $c=1$, and
$\lambda_k\simeq -\left(e^2/\pi R \right)\log k\ ,\ k\gg 1$.
A contribution to the confining potential is given by the Coulomb
interaction of the electrons
with the neutralizing ion background, which has also a droplet shape.
This determines \gps\ the phenomenological quantity
$\alpha_1\ =\ \left({e^2 /\pi}\right)\log \gamma R $, where $\gamma$ is
a non-universal parameter which depends on the details of
the droplet edge
\footnote{*}{
Note that the term proportional to $V^1_0$ in \hcoul\
does have the correct $1/R$ dependence anticipated
in \hren\ , but also a logarithmic correction, due to the Coulomb
infra-red singularity discussed in the previous section.}.
Other contributions to the confining potential can be included in
$\gamma$.

Let us now analyse the spectrum of  \hcoul\ for edge excitations
above the ground state $|\Omega\rangle$.
Since $V^0_0 = 0$ for all these excitations, it drops out of eq. \hcoul .
The Hamiltonian is diagonal in the bosonic basis \neut\ ,
because the Kac-Moody generators satisfy the ladder relations
$[H,V^0_{-k}]\propto V^0_{-k}\ $ (see eqs.\aff\Vir).
The c-number term in \hcoul,
\eqn\casi{
E_0\ = \ -{e^2\over 24 \pi R}\log \gamma R\ ,}
represents the finite-size Casimir energy
of the ground state \cdtz.

The fundamental excitations in the bosonic basis are the
one-boson states $V^0_{-k}|\Omega\rangle\ $,$ k > 0$, whose
energy is given by
\eqn\tos{
{\cal E}_k \ \equiv\ E_k -E_0\ =\
{e^2 \over \pi}\ {k \over R}\ \log \left({\gamma R\over k}\right)\ ,
\qquad 1\ll k\ll R\ .}
This spectrum has been previously found by
using various approximations: phenomenological classical hydrodynamics
\volkov \gps ,
edge bosonization \wep\ and a Feynman-like ansatz \gps.
Here, we gave a complete proof of this result,
free of unnecessary assumptions, in
the framework of the $\winf$ dynamical symmetry.
In the next section, we discuss how this spectrum has been confirmed
experimentally.

A virtue of our approach is the complete knowledge of the
Hilbert space of quantum edge excitations.
Given that we have found the form of the Hamiltonian
to leading order $O(1/R)$, we can find the complete spectrum of
all many-body states \neut.
Due to the ladder property of the operators $V^0_k$ with respect to the
Hamiltonian \hcoul\ ,
the energy of the many-boson states is simply additive in
the one-boson components,
\eqn\enpa{\eqalign{
H^{(Coulomb)}\ |\ k,\ \{ n_1, n_2,\dots , n_s \}\ \rangle\ &=\
{\cal E}_{\{n_1,...,n_s\}}\ |\ k,\ \{ n_1, n_2,\dots , n_s \}\ \rangle\ ,
\cr
{\cal E}_{\{n_1,...,n_s\}} \ &=\ \sum_{i=1}^{s}\ {\cal E}_{n_i}\ .\cr}}
This property of the many-body spectrum was first shown by
the numerical studies of Ref. \sto .
Actually, it is the characteristic property of the one-dimensional
bosonization of quantum edge excitations, which we have proved in this
paper,
and can be considered a success of the conformal field theory approach
to the quantum Hall effect.
Although the logarithmic deviation from the linear spectrum breaks scale
and conformal invariance, the conformal field theory techniques remain
still valid. In fact, the Hilbert space carrying the
representations of the Virasoro algebra (angular momentum)
also diagonalizes the non-conformal Hamiltonian \hcoul.
Actually, the $\winf$ algebra is richer than the Virasoro algebra,
because it includes more operators. These can be used to construct
non-conformal but still diagonalizable Hamiltonians.
This is one of the major points we would like to emphasize here.

\subsec{Short-range interactions}

It is also interesting to study the dynamics of edge excitations
in the presence of short-range interactions, which we have parametrized
with Gaussians
\eqn\spo{V({\bf x})\ =\ \int_{0}^{\infty}\ dt\ f(t)\
{\rm e}^{-t|{\bf x}|^2}\ ,}
and weight functions $f(t)$ of compact support centered around $t=1$.

We can add this interaction to the Hamiltonian \hcoul,
to describe short-distance deformations of the Coulomb interaction
due to microscopic effects, like the thickness of the layer.
Using the matrix elements of the short-range interaction
in eq.\lafin, we obtain the following contribution to the energy  of the
one-boson states \tos ,
\eqn\qcw{
\Delta {\cal E}_k\ =\ - {\pi \alpha \over{2 R^3}}\
k \left( k^2 -1 \right)\ , \qquad
\alpha\ =\ {1\over{4 \pi^{3/2} }}\ \int_{0}^{\infty} dt\ f(t)\
{ 1 + 2 t \over { \left( t(1+t) \right)^{3/2}} }\ . }

This spectrum
coincides with the classical capillary frequency \fdr,
for $\rho_0=eB\nu/(2\pi)=\nu/(\ell^2 \pi)$, and $\nu=1$,
in units $\ell = 1$.
It identifies the quantum expression for the edge tension
$\alpha$ as a ``moment'' of the interaction.
This result shows that the quantum
incompressible fluid behaves semiclassically
in the presence of short-range interactions, as anticipated in
the previous section.

Note that the result \qcw\ includes the Haldane short-range
interaction
\ref\hal{F. D. M. Haldane, {\it ``The Hierarchy of Fractional States
and Numerical Studies''}, in \prg .},\hfill\break
\hbox{ $H_I(|{\bf x}_1-{\bf x}_2|)=a\ \Delta
\delta^{(2)}({\bf x}_1- {\bf x}_2)$},
which has often been used in numerical and approximate analyses
to stabilize the $\nu=1/3$  Laughlin incompressible fluid.
In the present case, instead, it can be considered as a residual
interaction for edge excitations of the stable $\nu=1$ ground state.
The derivative of the delta function can be regularized by a Gaussian
with weight function
$f(t)= \lim_{t_o\to\infty}\ a\left( 8t^2/\pi\right)
\left( 2+ t\partial_t \right)\delta\left(t-t_o\right)$,
leading to the spectrum
\eqn\halda{
{\cal E}_k = -{a\over \pi\sqrt{\pi} R^3}\ k\left( k^2-1 \right)\ ,}
in agreement with previous results \sto\gps.

Another important property of eq. \qcw\ is that
the short-range contribution to the spectrum \tos\ is of
$O(1/R^3)$ and thus negligible for large $R$.
This proofs the {\it universality} of the dynamics of edge excitations:
in the thermodynamic limit, short-range interactions
become irrelevant for the dynamics of edge excitations, which
is only governed by the long-distance physics. The spectrum \tos\
yields further universal quantities, which add to
the kinematical data of section 2; these are combinations of
finite-size energies which are parameter-free (see also ref. \cdtz).

Finally, we would like to discuss the excitation spectrum in the case of
short-range interactions only. Since the experiments \weiss\
confirm the Coulomb spectrum \tos , this is a rather academic issue.
However, it allows us to clarify some aspects of the $\winf$ dynamics
associated with the higher spin operator $V^2_0$, which measures the
leading radial fluctuations of edge excitations. In this case,
we can consider the following Hamiltonian
\eqn\sronly{
H^{(sr)}\ =\ \alpha _0 V^0_0+ {\alpha _1\over R}
\left( V^1_0 -{1\over 24}\right) +
{\alpha _2\over R^2} \left( V^2_0-{1\over 12} V^0_0\right)\
+O\left({1 \over R^3}\right) .}
The term proportional to $V^2_0$ produces subleading effects which break
the leading scale-invariant spectrum of conformal field theory studied in
\cdtz.
The appropriate basis of edge excitations which diagonalizes $V^2_0$
(as well as all higher spin $\winf$ operators) is obtained from
the fermionic Fock operators,
\eqn\feb{
b^\dagger_{n_1} b_{m_1} \cdots b^\dagger_{n_s} b_{m_s} |\Omega\rangle\ ,
\qquad n_1>\cdots >n_s>0\ge m_1>\cdots >m_s\ ,
\quad \sum_i \left( n_i-m_i \right) =k\ .}
These are multiple particle-hole excitations with total angular
momentum $k$.
The operators $V^i_0$ are diagonal in this basis because the
pair $\left( b^\dagger_n b_m \right)$ is a ladder operator for them
(see eqs. \acf\vofs). The spectrum of one particle-hole excitation
$b^\dagger_nb_m |\Omega\rangle$ follows from \vofs :
\eqn\cosa{
{\cal E}_{nm}=E_{nm}-E_0 =
{\alpha_1 \over R} \left( n\ -\ m \right)\ +\ {\alpha _2 \over{R^2}}
\left( n- m\right)\left(n+m-1 \right)\ .}
Since the indices $(n,m)$ can be traced back to the ones of
$w_\infty$ classical deformations $\delta_{nm}\rho$ in \wav, we find that
quantum radial fluctuations indeed break the classical degeneracy at
fixed $k=(n-m)$. Note that this degeneracy is still present in the
leading conformal spectrum. For a given value of $k$,
the non-degenerate eigenvalues are distributed in the interval
\eqn\inte{
{\alpha_1 \over R} k\ -\  {\alpha _2 \over{R^2}} k(k-1)
\ \le\ {\cal E}_{nm}\ \le
{\alpha_1 \over R} k\ +\  {\alpha _2 \over{R^2}} k(k-1)\ .}
Note, however, that the breaking of conformal invariance generated
by $V^2_0$ is different
from the one induced by the Coulomb interaction, because the fermionic
and bosonic bases are different.
The fermionic splitting of the conformal spectrum is characteristic
of another $(1+1)$-dimensional model with $\winf$ symmetry:
the Calogero-Sutherland model
\ref\calo{ F. Calogero, {\it J. Math. Phys.} {\bf 10} (1969) 2197;
{\it ibid.} {\bf 12} (1971) 418;
B. Sutherland, {\it J. Math. Phys.} {\bf 12} (1971) 246;
{\it Phys. Rev.} {\bf A 4} (1971) 2019; {\it ibid.} {\bf A 5}
(1972) 1372.} .
In this case, it has been show
\ref\polk{A. P. Polychronakos, {\it Phys. Rev. Lett.} {\bf 69} (1992)
703; D. V. Khveschenko, preprint cond-mat/9401012.}
that the
two-body interaction can be written entirely in terms of the
generator $V^2_0$. In the
quantum Hall effect, instead, the interaction cannot be written in terms
of the generators $V^i_0$ only.

\subsec{Charged excitations}

We can easily extend the previous discussion to compute the spectrum
of charged edge excitations. Let us consider states with $q$
electrons ($q\in {\rm Z}$) added to the boundary and,
correspondingly, $q$ vortices in the
interior of the droplet. These are the highest-weight
states $|Q\rangle$ for other $\winf$ representations,
with weights given by \hws.
In particular, the weights of $V^0_0$ and $V^1_0$ are given by
$Q=q$ and $J=q^2/2$, respectively, and determine
the normal-ordered energy of these states by eq. \hcoul,
\eqn\eqq{
H|Q=q\rangle\ =\ E_{0,q}|Q\rangle\ ,\qquad
E_{0,q}=\alpha_0 q + {e^2\over \pi R}\left( {q^2\over 2} -{1\over 24} \right)
\log\gamma R\ + \ {e^2\over \pi R} q^2 \ .}
Clearly, this is only a contribution to the energy gap of these states.
The main contribution comes from the physics at the core of the
vortices in the deep interior of the droplet, which cannot be computed
in the boundary theory \lau.
However, eq. \eqq\ gives the universal part of this gap \cdtz.

The edge excitations on top of both the charged and neutral
states have the same structure.
The energy levels for $Q\neq 0$ are only shifted by a $Q$-dependent
constant. In particular, the Coulomb interaction yields the spectrum \tos\
for the energy differences $(E_{k,q} -E_{0,q})$.

\subsec{Filling $\nu=1/(2p+1)$}

We can use the $\winf $ dynamical symmetry to generalize the
previous results for the spectrum of edge excitations  to the
Laughlin fluids at fractional filling $\nu=1/3,1/5,1/7,\cdots$.
As discussed in section 2.2, these excitations
are given by other representations of the
$\winf$ algebra with central charge $c=1$ \cdtz\ctzb
\ref\flohr{M. Flohr and R. Varnhagen, {\it J. Phys.} {\bf A 27} (1994)
3999; D. Karabali, {\it Nucl. Phys.} {\bf B 419} (1994) 437.}.
In reference \cdtz, we have shown that these representations have
one free parameter, which is fixed by the value of the
Hall conductivity.
The operators $V^0_k$, normalized to measure the physical charge,
satisfy the Kac-Moody algebra,
\eqn\trickkm{
[\ V^0_n, V^0_m\ ] = {n\over 2p+1}\ \delta_{n+m,0}\ ,
\qquad \left(\nu={1\over 2p+1}\ , p=1,2,\dots\ \right) \ ,}
and the spectrum of anyon highest weights is given by
$Q=q/(2p+1)$, $J=q^2/(4p+2)$, with $q\in {\rm Z}$ \cdtz\ctzb.

Lacking a microscopical derivation of the Coulomb Hamiltonian for
fractional filling, we shall use an argument based on the $\winf$
dynamical symmetry. Namely, we shall argue that the Hamiltonian for edge
excitations must be {\it form invariant}, {\it i.e.}, must
take the same form
\footnote{*}{Note that this corresponds to a complicated
microscopic expression, which incorporates the idea
of ``attaching flux tubes to electrons'' \jain , see also \flohr .}
\hcoul\ in terms of the $\winf$ operators
for any value of  $\nu=1/(2p+1)$. Indeed,
the structure of edge excitations, {\it i.e.}, of the $\winf$
highest-weight representations, is identical for all these values
of $\nu$.
Therefore, we can repeat the analysis after \hcoul\ , and use the
commutators \trickkm\ and \Vir\ to compute the corresponding spectrum
of one-boson neutral edge excitations,
\eqn\tosnu{
{\cal E}_k \ =\ E_k -E_0\ =\
{e^2 \nu\over \pi}\ {k \over R}\ \log \left({\eta R\over k}\right)\ ,
\qquad 1\ll k\ll R\ ,\quad\left(\nu={1\over 2p+1}\right)\ ,}
where the parameter $\eta$ may depend on $\nu$.
This result is again semiclassical \wep\gps.
It is possible, although more difficult, to
generalize our results to the complete hierarchy of incompressible
quantum fluids corresponding to $\winf$ theories with $c > 1$ \ctzb.


\newsec{Conclusions}

In this paper, we further developed the $\winf$ effective field theory
for the plateaus of the quantum Hall effect.

We first reviewed how the $\winf$
representations describe the kinematical data of the Laughlin
incompressible quantum fluids, like the fractional charges and
statistics of anyon excitations.
Secondly, we expressed the general Hamiltonian for edge excitations
in terms of the $\winf$ generators for both cases of Coulomb and
short-range interactions, and obtained its spectrum.
We proved the one-dimensional bosonization of edge excitations
by showing that the leading term in the $1/R$ expansion of the
Hamiltonian contains the Kac-Moody and Virasoro operators
only. Moreover, for large $R$, the spectrum is independent of
short-range effects and thus possesses universal features.

Furthermore, we showed that the infinite tower of higher
spin operators $V^i_k,\ i \ge 2\ $,
parametrizes the subleading corrections to the bosonization,
which arise from radial fluctuations of the edge excitations.
This is the physical picture for the ``$(2+1)$-dimensional
nature'' of the $\winf$ algebra recently remarked in Ref. \awat .

Finally, we briefly discuss the experiment \weiss\ of radio-frequency
resonance,
which confirmed the finite size spectra \tos\ and \tosnu, for the case of
the Coulomb interaction. Low dissipation excitations
(called edge magnetoplasmons there) were observed as sharp
resonance peaks in the transmitted signal, at the plateaus for
$\nu=1, 2$ and $2/3$. The observed
resonance frequencies $\omega_i = {\cal E}_i /\hbar$ match the
formulae \tos\ and \tosnu , and verify both their logarithmic functional
form and their $\nu$ dependence
(see table I and eq. (7) of ref. \weiss ).
Let us also quote the recent experiment with very precise time
resolution \ash ,
which clearly establishes the one-dimensional and chiral nature
of the edge excitations.
\bigskip
\bigskip
{\bf Acknowledgements}
\bigskip
We would like to thank G. V. Dunne, F. D. M. Haldane, M. Stone
and S. Stringari for interesting discussions and S. Kivelson
for calling to our attention ref. \weiss . We also thank
CERN and the Department of Physics and INFN of Florence for
their kind hospitality.

\vfill
\eject

\appendix{A}
\bigskip
\noindent{\bf $\winf$ algebra and Weyl fermion representation}
\bigskip
In this appendix, we derive explicit expressions for the $\winf$ algebra,
the form of the
generators $(V_R)^i_n$ defined on the cylinder geometry, and the
realization of the $\winf$ algebra in terms of a Weyl fermion.
\bigskip
\bigskip
\noindent{\it $\winf$ algebra}
\bigskip
The $\winf$ algebra can be written in a compact first-quantized form
\kac,
\eqn\winfalg{\eqalign{
[V\left( z^r f(D) \right) ,V\left( z^s g(D) \right)]=
&V\left( z^{r+s}f(D+s)g(D) \right) - V\left(z^{r+s}f(D)g(D+r) \right) \cr
&+ c\ \Psi \left( z^r f(D), z^s g(D) \right) \ ,\cr} }
where the central extension is
\eqn\cen{
\Psi\left( z^r f(D), z^s g(D) \right) \equiv\
\delta_{r+s,0}\ \sum_{j=1}^r\ f(-j)g(r-j)\ \equiv
- \Psi\left( z^s g(D), z^r f(D) \right)\ , \quad  r>0\ .}
In these expressions, the $\winf$ operators $V\left( z^r f(D) \right)$
are specified by the corresponding first-quantized differential
expression $z^r f(D)$, where $f(D)$ is an entire function of
$D \equiv z {\partial \over{\partial z}}$. In the text, we introduced
the operators $V^i_k$, characterized by mode
index $k\in{\rm Z}$ and conformal spin $h=i +1 \ge 1$: the two
definitions are related by
\eqn\defw{
V^i_k \equiv V\left( z^k f^i_k(D) \right) \ ,}
where $f^i_k(D)$ are specific $i$-th order polynomials.
The basis for these polynomials is chosen by requiring that the central
extension \cen\ is diagonal in the spin indices. This can be done
uniquely, because $\Psi$ has the property of a symplectic metric.
The generators of lower spin are thus found to be:
\eqn\lowcurr{\eqalign{
V^0_n = & V\left( -z^n \right) \ , \cr
V^1_n = & V\left( -z^n \left( D+{1\over 2}(n+1) \right) \right) \ , \cr
V^2_n = & V\left( -z^n \left( D^2 +(n+1) D +{1\over 6}(n+1)(n+2)\right)
          \right) \ , \cr
V^3_n = & V\left( -z^n \left( D^3 +{3\over 2}(n+1) D^2 +
        {6n^2 +15n+11\over 10} D +{(n+3)(n+2)(n+1)\over 20} \right)
          \right) \ , \cr} }
where $V^0_k$ and $V^1_k$ are, by definition, the Kac-Moody and
Virasoro generators, respectively \gin.
Some of the algebraic relations are:
\eqn\com{\eqalign{
\left[ V^0_n, V^0_m \right] &= n\ c\ \delta_{n+m,0}\ ,\cr
\left[ V^1_n, V^0_m \right] &= -m\ V^0_{n+m}\ ,\cr
\left[ V^1_n, V^1_m \right] &= (n-m)\ V^1_{n+m} + {c\over 12}\left(
n^3-n\right) \delta _{n+m,0}\ ,\cr
\left[ V^2_n, V^0_m \right] &= -2m\ V^1_{n+m}\ ,\cr
\left[ V^2_n, V^1_m \right] &= (n-2m)\ V^2_{n+m} -
     {1\over 6}\left(m^3-m\right) V^0_{n+m}\ ,\cr
\left[ V^2_n, V^2_m \right] &= (2n-2m)\ V^3_{n+m}
     +{n-m\over 15}\left( 2n^2 +2m^2 -nm-8 \right) V^1_{n+m} \cr
     &\quad +\ c\ {n(n^2-1)(n^2-4)\over 180}\ \delta_{n+m,0}\ .\cr} }

Any highest weight representation is formed by all the states obtained
by applying $\winf$ generators with negative moding
to the highest weight state $|Q\rangle$,
which satisfies,
\eqn\hwcond{
V^i_n|Q\rangle = 0\ , \qquad \forall\ n>0\ ,\ i\ge 0 \ ,}
and is the eigenstate of $V^i_0$,
\eqn\hweige{
V^i_0|Q\rangle = m_i(Q)|Q\rangle\ .  }
We shall only discuss the unitary, irreducible representations for $c=1$,
which describe the Laughlin incompressible fluids with
$\nu=1/(2p+1)\ $, $p=0,1,2,\cdots$ \cdtz. For these representations,
the $m_i(Q)$ are $(i+1)$-th polynomials in $Q$, which are given by the
generating function,
\eqn\eigenfu{
V\left( -{\rm e}^{xD} \right) |Q\rangle =
   {{\rm e}^{xQ} - 1 \over {\rm e}^x -1}\ |Q\rangle \ .}
For example, by expanding in powers of $x$, and using \lowcurr, one finds
$m_0(Q) =Q\ $ and $\ m_1(Q)=J=Q^2/2\ $.
The eigenvalues of $V^0_0/(2p+1)$  and $V^1_0$ measure the physical
charge and the angular momentum of the states, respectively \cdtz\ctzb.
For $c=1$, all the states in a unitary irreducible representation can
obtained by using the Kac-Moody generators only \kac.
These states have the form,
\eqn\verma{
|k,\{n_1,n_2,\dots,n_s\}\rangle =
V^0_{-n_1}V^0_{-n_2}\cdots V^0_{-n_s}|Q\rangle
\ ,\qquad n_1\ge n_2 \ge \cdots \ge n_s >0\ ,}
where $k=Q^2/2 + \sum_{i=1}^s n_i\ $ is  their angular momentum.
\bigskip
\bigskip
\noindent{\it $\winf$ generators on the cylinder}
\bigskip
In the quantum Hall effect, fields and operators are
naturally defined on the boundary
circle $(0\le\theta <2\pi)$, {\it i.e.},
on a compact space.
In the mathematical literature, however, they are
conventionally considered in an unbounded space.
There is a conformal mapping between these two spaces,
which are the {\it cylinder}
$(u=\tau -iR\theta)$ and the {\it conformal plane} $(z)$,
after inclusion of the euclidean time $\tau$,
\eqn\cft{
z=\exp\left({u\over R}\right)=
\exp\left({\tau\over R} -i\theta \right) \ .}
For example,
the operators \lowcurr\ define the $\winf$ currents $V^i(z)$
on the conformal plane as follows,
\eqn\hsc{
V^i(z)\ \equiv\ \sum_n\ V^i_n\ z^{-n-i-1}\ .}

On the other hand, the Hamiltonian of edge excitations should be
expressed in terms of the $\winf$ generators $(V_R)^i_n$
on the cylinder. The latter can be obtained by transforming the
$V^i(z)$ under the conformal mapping \cft, as follows.
The generator $G$
of infinitesimal conformal transformations $z=u +\epsilon(u)$,
with $\epsilon(u)= \sum_n \epsilon_n u^{n+1}$, is parametrized
by the Virasoro operators,
\eqn\defg{G\ =\ \sum_n\  \epsilon_n\ V^1_n\ .}
Therefore, the infinitesimal transformation of the $\winf$ currents,
$V^i(u)= V^i(z) + \delta_{\epsilon} V^i(z)$, are computed from
the algebra \com,
\eqn\itc{\eqalign{
\delta_{\epsilon} V^0\ &\equiv\ [ G, V^0 ]=\epsilon \partial
V^0 + (\partial\epsilon) V^0\ , \cr
\delta_{\epsilon} V^1\ &\equiv\ [ G, V^1 ]=\epsilon \partial
V^1 + (\partial\epsilon) V^1+ {c\over{12}} \partial^3 \epsilon\ , \cr
\delta_{\epsilon} V^2\ &\equiv\ [ G, V^2 ]=\epsilon \partial
V^2 + (\partial\epsilon) V^2+{1\over 6}(\partial^3 \epsilon)V^0
\ . \cr}}
Note that these transformations contain the differential operator
$[\epsilon\partial + h(\partial\epsilon)]$ applied to the current,
plus additional terms. If the latter are absent, as in the case of
$V^0(z)$, the infinitesimal transformation can be integrated to
yield the finite form
\eqn\cftmap{\Phi (u)\ =\ \left({dz\over{du}} \right)^h\
\Phi(z(u))\ .}
Those fields $\Phi(z)$ which transform homogeneously are called
{\it primary} \gin.
The current $V^i$, $i >  0$, have additional terms which, nevertheless,
vanish for $\epsilon=\epsilon_{-1}+\epsilon_0 z+ \epsilon_1 z^2$,
{\it i.e.}, for global conformal transformations generated
by $(V^1_{-1},V^{1}_0,V^1_1)$.
There is only one function $S(z(u),u)$ of the conformal mapping, which
vanishes under the finite global transformations: the Schwartzian
derivative \gin
\eqn\swd{S(f(z),z)\ \equiv\ {f^{\prime\prime\prime} \over{f^{\prime}}}
- {3\over2}\left({f^{\prime\prime}\over{f^{\prime}}}\right)^2 \ ,}
where the prime denotes differentiation. Owing to this property,
all infinitesimal transformations in \itc\ can be integrated, and
yield
\eqn\ftr{\eqalign{
V^0 (u)\ &=\ {dz\over{du}}\ V^0(z)\ , \cr
V^1 (u)\ &=\ \left({dz\over{du}}\right)^2\ V^1(z)\ + {c\over12}\
S(z,u)\ , \cr
V^2 (u)\ &=\ \left({dz\over{du}}\right)^3\ V^2(z)\ + {1\over 6}\
{dz\over{du}}\ S(z,u)\ V^0(z)\ . \cr }}
The $\winf$ currents on the cylinder are thus found by using
the mapping \cft,
\eqn\wcurc{\eqalign{
V^0_R (u)\ &= {z\over R} V^0(z)\ ,\cr
V^1_R (u)\ &= {1\over R^2}\left( z^2 V^1(z) - {1\over 24}\right)\ ,\cr
V^2_R (u)\ &= {1\over R^3}\left( z^3 V^2(z) -
{z\over 12}V^0(z)\right)\ .\cr}}
Using the definition
\eqn\dhw{
 \left(V_R\right)^j_0\equiv\ \int _0^{2\pi iR} { du \over{-2\pi i}}
\ V^j_R (u)\ ,}
we obtain the relation between the zero modes in the two geometries
\eqn\wgc{
\left(V_R\right)^0_0 = V^0_0\ ,\qquad
\left(V_R\right)^1_0 = {1\over R} \left( V^1_0-{c\over 24} \right) \ ,
\qquad
\left(V_R\right)^2_0 = {1\over R^2} \left( V^2_0
-{1\over 12} V^0_0 \right) \ .}
\bigskip
\bigskip
\noindent{\it Weyl fermion representation of $\winf$}
\bigskip
In the thermodynamic limit \cdtz, the field operator of electrons
in the lowest
Landau level can be evaluated at the boundary of the droplet,
and mapped into the $(1+1)$-dimensional field of the chiral,
relativistic Weyl fermion \gin,
\eqn\weyf{
F_R(\theta) ={1\over\sqrt{ R}}\ \sum_{k=-\infty}^{\infty}
{\rm e}^{i\left( k-{1\over 2} \right) \theta}\ b_k \ ,}
with Neveu-Schwarz boundary conditions on the circle
(the $b_k$ are the fermionic Fock operators).
The vacuum of this theory corresponds to the ground state of the $\nu=1$
incompressible fluid $|\Omega\rangle$, which satisfies,
\eqn\vachwc{
b_l|\Omega\rangle =0 \ ,\quad l>0\ ,
\qquad b^\dagger_l|\Omega\rangle=0\ ,
\quad l\le 0\ .}
The Hilbert space of a Weyl fermion consists of an infinity of
$c=1$ $\winf$ representations \hwcond - \verma, all
those corresponding to $Q \in{\rm Z}$. For example, $|\Omega\rangle$ is
the highest weight state \hwcond\ with $Q=0$.

The Weyl fermion is a primary conformal field \cftmap\ with $h=1/2$,
and on the plane $z$ takes the form $(\tau =0)$,
\eqn\weypl{
F(z)= \left({du\over dz}\right)^{1/2}  F_R(\theta)=
\sum_l \ {\rm e}^{il\theta}\ b_l \ , \qquad
{\bar F}(z)= \left({du\over dz}\right)^{1/2}  F^\dagger_R(\theta)=
\sum_l \ {\rm e}^{-i(l-1)\theta}\ b^\dagger_l \ . }

The representation of $\winf$ operators on this Hilbert space is
obtained by sandwitching the first-quantized expressions \defw\
between the field operators,
\eqn\defwf{
V^i_n \equiv \oint \ {dz\over 2\pi i}\ : F(\theta)\
z^n\ f^i_n(D) \ {\bar F}(\theta):\  =
\oint \ {dz\over 2\pi i}\ : {\bar F}(\theta)\
z^n\ g^i_n(D) \ F(\theta):\ ,}
where $g^i_n=(-1)^{i+1}f^i_n$, and
the integration is carried clockwise over the unit circle.
Since the anticommutator of fields is a delta function in Fock
space, the $V^i_k$ defined above clearly represent the algebra \winfalg.
One can verify that they also have the eigenvalues \eigenfu\ when acting on
$Q$ fermion states.
In eq. \defwf, they are also written in the canonical form $({\bar F}F)$
of quantum field theory, which identifies the polynomials
$g^i_n$ defined in section 2.2.
Moreover, the normal ordering prescription $\ :\ :\ $ w.r.t. the ground
state \vachwc\  is obtained by writing the annihilation operators
$b_l \ (l>0)$ and $b^\dagger_l\ (l\le 0)$ to the right-hand side of the
creation operators $b_l \ (l\le 0)$ and $b^\dagger_l\ (l> 0)$.

{}From the expressions \lowcurr , we obtain the Fock space expression
of the generators,
\eqn\fockw{\eqalign{
V^0_k =&\ \sum_l\ :  b^\dagger_{l-k}\ b_l :\ ,\cr
V^1_k =&\ \sum_l\left(l-{k+1\over 2}\right)\
:  b^\dagger_{l-k}\ b_l :\ ,\cr
V^2_k =&\ \sum_l\left(l^2 -(k+1)l + {{(k+1)(k+2)}\over 6}
\right) :  b^\dagger_{l-k}\ b_l :\ ,\cr}}
and the form of the currents \hsc\ on the plane,
\eqn\wcurp{\eqalign{
V^0(z)=& :{\bar F}F: \ ,\cr
V^1(z)=& {1\over 2}:\partial_z{\bar F}F:
- {1\over 2}:{\bar F}\partial_z F: \
,\cr
V^2(z)=& - :\partial_z{\bar F}\partial_z F: +{1\over 6}:\partial_z^2
        \left({\bar F}F\right): \ .\cr}}

Let us finally remark that the
different form \wcurc\ of the $\winf$ currents on the plane and
the cylinder is due to a normal ordering effect \car.
Actually, we can recover
\wcurc\ in the fermionic case ($c=1$) as follows.
We apply the conformal mapping \cft\ to each fermion field
in \wcurp, paying attention to the different normal ordering in the plane
and the cylinder. For example,
\eqn\exp{\eqalign{
V^0_R (u)=&:{\bar F}_R (u)F_R (u):\equiv\lim_{u_1,u_2\to u}
         \left({\bar F}_R (u_1) F_R (u_2) - {1\over u_1-u_2} \right) \cr
      =& {dz\over du}\ V^0(z) + \lim_{u_1,u_2\to u}\left[
        \left({dz_1\over du_1}{dz_2\over du_2}\right)^{1/2}
         {1\over z_1-z_2} -{1\over u_1-u_2} \right]
      =  {z\over R}\ V^0(z) \ .\cr}}
Proceeding similarly for the other currents, we indeed recover \wcurc.

\vfill
\eject

\appendix{B}
\bigskip
\noindent
{\bf The Gaussian interaction two-body matrix element when $N\to\infty$}
\bigskip
In this appendix, we evaluate the two-body matrix element for the
Gaussian interaction \mesr\ in the limit $N\to\infty$. It is
is given by the expression
\eqn\apa{\eqalign{M(N+r,N+s;k)\ = {1\over 2} &\int {d^{2} z_1 d^2 z_2
\over{\pi^2}}\ {\rm e}^{ -|z_1 |^2 -|z_2 |^2}\
{ {\bar z}^{N+s}_1 z^{N+s+k}_1 \over{\sqrt{ (N+s)! (N+s+k)!}} }
{\rm e}^{-t|z_1 - z_2|^2}\ \cdot \cr
&{ {\bar z}^{N+r}_2 z^{N+r-k}_2 \over{\sqrt{ (N+r)! (N+r-k)!}} }\ . \cr}}
The idea is to evaluate exactly the angular integrations in \apa\ and
use the saddle-point technique to calculate the integrals that involve
the modulus of the coordinates, when $N\to\infty$. We first
rewrite \apa\ upon replacing $z_1= x{\rm e}^{i\theta}$, $z_2 = y$,
with $x,y \ge 0$:
\eqn\apb{\eqalign{ M(N+r,N+s;k)\ &=\ {2 \over{C(N,s,r,k)}}
\int_{0}^{\infty} dx\ x^{2N+2s+k+1}
\int_{0}^{\infty} dy\ y^{2N+2r-k+1}\ \cdot \cr
&\qquad {\rm e}^{-(1+t)(x^2 + y^2)}
I_k (2txy)\ , \cr
C(N,s,r,k)\ &\equiv\ \sqrt{ (N+s)! (N+s+k)! (N+r)! (N+r-k)! }\ ,\cr
I_k (z)\ &=\ {1\over{2\pi}} \int_{0}^{2\pi} d\theta\ {\rm e}^{ik\theta
+ z \cos \theta}\ .\cr}}
For $N\to\infty$, the most important contribution to the radial
integrations in \apb\ arise from values of $x,y \sim O(\sqrt{N})$.
Therefore, for values of $t$ that are not too small, one can
assume that the argument of the imaginary Bessel function $I_k(z)$ in
\apb\ satisfies $z=2txy \gg 1$. In this regime, one may use the
asymptotic form \gradst :
\eqn\apc{ I_k(z)\ \simeq\ {{\rm e}^z \over{\sqrt{2\pi z}}} \left(
1\ -\ {1\over{2z}}(k^2 - 1/4)\ + O\left( {1\over z^2} \right)
\right)\ ,\qquad z=2txy \gg 1\ .}
Replacing \apc\ in \apb\ , one obtains
\eqn\apd{\eqalign{ M(N+r,N+s;k)\ &=\ {1 \over{C(N,s,r,k)}}
{1\over{\sqrt{\pi t}}}\int_{0}^{\infty}
\int_{0}^{\infty} dx\ dy\ {\rm e}^{\Phi(x,y)}\ ,\cr
\Phi(x,y)\ &\equiv -(1+t)(x^2 + y^2) + 2txy + (2N+2s+k+1/2)\ln x \cr
&+ (2N+2r-k+1/2)\ln y\ .\cr}}
One then finds the two-dimensional saddle-point of \apd\ $(x_0 , y_0)$:
\eqn\ape{\eqalign{ x^2_0\ &=\ N+1/4 +{{2s + k +t(r+s)}\over{2(1+t)}}
+O \left( 1/N \right)\ ,\cr
y^2_0\ &=\ N+1/4 +{{2r - k +t(r+s)}\over{2(1+t)}}
+O \left( 1/N \right)\ .\cr}}
Therefore, the saddle-point value of \apd\ is given by:
\eqn\apf{\eqalign{
M(N+r,&N+s;k)\ =\ {1 \over{C(N,s,r,k)}} {1\over{\sqrt{\pi t}}}\ \cdot \cr
 &\left\{ {\rm e}^{\Phi(x_0 ,y_0 )}\ {2\pi}\
\left ( \det \left(
{\partial^2 \Phi \over{\partial x \partial y}} \right)_0 \right)^{-1/2}
- {(k^2 - 1/4)\over{4t}}\left({r\to r-1/2  \atop  s \to s-1/2}
\right)\right\} .\cr}}
The value of the Jacobian is given by:
\eqn\jac{\det \left( {\partial^2 \Phi \over{\partial x \partial y}}
\right)_0 = 16(1+t) + O\left( 1/N^2 \right)\ .}
Replacing now \ape\ and \jac\ into \apf , and retaining terms up to
$O(1/N^{3/2})$, one finds
\eqn\apg{\eqalign{M(N+r,&N+s;k)\ = {1\over{\sqrt N}} {1\over{4 \sqrt{\pi}
\sqrt{t(1+t)}} } \left( 1- { k^2 - 1/4 \over{4t N}} \right) \cdot \cr
&\left\{1 + {1\over{2N}} \left( k (r-s-k) - r^2 - s^2 - {r+s \over 2}
- {5\over 24} \right)\right. \cr
&+ \left. {k^2 -2k(r-s) +2 (r^2 + s^2) + t(r+s)^2
\over {4N (1+t)} }\right\}\ . \cr}}
which is formula \meres .
\vfill

\listrefs
\end